\documentclass[english]{paper}
\usepackage[T1]{fontenc}
\usepackage[latin9]{inputenc}
\usepackage[letterpaper]{geometry}
\usepackage{amsmath}
\usepackage{amssymb}
\usepackage[numbers]{natbib}

\makeatletter
\usepackage{xcolor}
\usepackage[
  colorlinks=true,
  citecolor=red,   
  linkcolor=red,    
  urlcolor=black
]{hyperref}

\makeatother

\usepackage{babel}
\begin{document}
\title{Coherent-field QED transitions based on coherent-state boundary conditions}
\author{Keita Seto$^{1,2}$\thanks{email: seto.keita@nifs.ac.jp}}
\institution{$^{1}$ National Institute for Fusion Science, National Institutes of Natural Sciences,\\
    322-6 Oroshi-cho, Toki, Gifu, 509-5292, Japan\\
\\
$^{2}$ The Graduate University for Advanced Studies, SOKENDAI\\
    322-6 Oroshi-cho, Toki, Gifu, 509-5292, Japan}

\maketitle
\noindent \today
\begin{abstract}
We develop a coherent-field formulation of quantum electrodynamics
(QED) based on coherent-state boundary conditions, in which laser
fields are represented by asymptotic coherent states rather than by
prescribed classical background fields. Starting from coherent-state
boundary conditions and the displacement-operator formalism, we construct
operator and path-integral descriptions of transitions between distinct
electromagnetic coherent states. This formulation provides a coherent-field
extension of conventional background-field QED and incorporates the
quantum dynamics of the coherent field itself. The corresponding path-integral
representation contains a functional integral over coherent-field
histories in addition to the usual functional integral over quantum
fluctuations, thereby extending the conventional background-field
description. 

The resulting path-integral representation naturally leads to an effective
action for the stationary coherent-field configuration. The corresponding
saddle-point condition yields an effective Maxwell equation containing
contributions from vacuum polarization, photon fluctuations, and coherent-field
fluctuations. In the limit where the latter two contributions vanish,
the formalism reduces to the conventional Heisenberg\textendash Euler
description.

We further derive a saddle-point expression for the transition probability
between coherent-field sectors, $P_{\alpha\rightarrow\alpha^{\prime}}$.
For identical initial and final coherent states, $P_{\alpha\rightarrow\alpha}$
may be interpreted as the survival probability of a coherent-field
sector, while its complement provides a useful indicator of the tendency
to leave the original coherent-field sector. As a first qualitative
application, we examine the weak coherent-field limit using the leading-order
Heisenberg\textendash Euler effective action and show that the depletion
indicator depends not only on the imaginary part of the effective
action but also on field-dependent modifications of the fluctuation
spectrum.

The present formulation clarifies the relation between coherent-field
QED and conventional background-field QED, and provides a framework
in which transitions between distinct coherent-field sectors are treated
as fundamental quantum processes.
\end{abstract}

\section{Introduction}

Quantum electrodynamics (QED) predicts that the vacuum behaves as
a nonlinear medium in the presence of sufficiently strong electromagnetic
(EM) fields \citep{Heisenberg-Euler1936,Schwinger1951,Greiner_Muller_Rafelski_1985,Fradkin-Gitman-Shvartsman1991,Marklund2006}.
Representative examples include vacuum polarization described by the
Heisenberg\textendash Euler effective action \citep{Heisenberg-Euler1936,Euler_Kockel1935,Weisskopf1936}
and vacuum instability associated with the Schwinger mechanism \citep{Schwinger1951,Sauter1931,Nikishov1970}.
The interpretation of the corresponding vacuum-decay probability and
its relation to pair-production observables has been discussed in
detail in Ref.\,\citep{Cohen_McGady_2008}. Vacuum polarization gives
rise to a variety of nonlinear optical phenomena, including vacuum
birefringence, photon-photon scattering, and modified EM-wave dispersion
relations \citep{Marklund2006,Euler_Kockel1935,Dittrich-Gies2000}.
These phenomena are commonly formulated within the conventional background-field
approximation, or Furry-picture framework, in which the EM background
is treated as a prescribed classical field while the quantum dynamics
of electrons, positrons, and photons are considered on top of that
background \citep{Volkov1936,Furry1951,Ritus1985,Piazza2012}.

This strong-field QED framework \citep{Ritus1985,Piazza2012} underlies
the Furry-picture formulation \citep{Furry1951} and the description
of particle dynamics in prescribed laser backgrounds through Volkov
states \citep{Volkov1936}. Nevertheless, this fixed-background approximation
possesses an intrinsic limitation: the background field itself is
assumed to remain fixed throughout the quantum evolution. As a result,
effects associated with the modification of the laser field, such
as depletion, backreaction, or changes in its coherent-state structure,
cannot be described within the conventional formulation. This limitation
has motivated several attempts to go beyond the fixed-background approximation,
including dynamical treatments of laser depletion and backreaction
in strong-field QED \citep{Ilderton-Seipt2018,Adamo-Ilderton}.

This limitation becomes increasingly relevant in the context of ultra-intense
laser systems. In the standard vacuum-persistence approach of QED
with a background laser EM field, the quantity 
\begin{equation}
P_{\mathrm{vacuum}}(\mathcal{A})\propto|\langle0_{(\mathrm{out})}(\mathcal{A})|0_{(\mathrm{in})}(\mathcal{A})\rangle|^{2}
\end{equation}
plays a central role in the conventional fixed-background formulation
of strong-field QED for a prescribed EM background $\mathcal{A}$,
and the probability loss 
\begin{equation}
1-P_{\mathrm{vacuum}}(\mathcal{A})
\end{equation}
is commonly interpreted as a measure of vacuum decay and particle
production \citep{Schwinger1951,Greiner_Muller_Rafelski_1985,Fradkin-Gitman-Shvartsman1991,Dunne2005}.
However, because the background field is treated as fixed from the
outset, this quantity does not resolve whether the reduction of the
vacuum-persistence weight originates from real particle-production
channels or from changes in the EM field configuration itself. In
other words, the conventional formalism does not provide access to
transitions between different EM-field configurations such as $\langle0_{(\mathrm{out})}(\mathcal{A}^{\prime})|0_{(\mathrm{in})}(\mathcal{A})\rangle$.

Recent studies of depletion and backreaction effects in strong-field
QED have emphasized that an intense laser pulse should ultimately
be regarded as a quantum state of the EM field rather than as a strictly
fixed classical background \citep{Ilderton-Seipt2018,Adamo-Ilderton}.
From this viewpoint, the laser is naturally described by a coherent
state, which provides the closest quantum counterpart of a classical
EM wave and allows the background EM field to be treated as a quantum
state rather than as a prescribed external input. Coherent-state techniques
have long been established in quantum optics and many-body physics
\citep{Klauder-Skagerstam_1985,Negele-Orland1988}, and coherent states
play a central role in the quantum description of laser radiation
\citep{Glauber1963a,Glauber1963b,Sudarshan1963,Sargent1974,Loudon_2000}.
Nevertheless, a formulation in which coherent-state boundary conditions
are incorporated directly into the QED effective-action framework
remains largely unexplored.

The purpose of the present work is to formulate QED directly in terms
of coherent-state boundary conditions and to investigate transitions
between distinct EM coherent fields. Starting from the Gupta\textendash Bleuler
condition \citep{Gupta1950,Bleuler1950}, we construct a coherent-state
representation of laser fields \citep{Glauber1963a,Glauber1963b,Sudarshan1963,Sargent1974,Loudon_2000}
and reformulate the corresponding QED both in operator language and
in the path-integral formalism. The background EM field appears naturally
as the expectation value of a coherent state rather than as an externally
prescribed classical field.

Within this framework, transition amplitudes between different coherent
states, 
\begin{equation}
|\alpha_{(\mathrm{in})}\rangle\rightarrow|\alpha_{(\mathrm{out})}^{\prime}\rangle,
\end{equation}
are described by a coherent-field path-integral representation obtained
by translating the standard coherent-state path integral \citep{Klauder-Skagerstam_1985,Negele-Orland1988}
into a functional integral over the corresponding coherent-field histories.
The resulting effective action combines this coherent-field path-integral
representation with the effective-action framework of QED \citep{Heisenberg-Euler1936,Schwinger1951}
and incorporates vacuum-polarization effects, photon fluctuations,
and coherent-field fluctuations arising from the functional integral
over coherent-field histories. The corresponding stationary condition
yields an effective Maxwell equation containing fluctuation-induced
currents beyond the conventional Heisenberg\textendash Euler description
\citep{Heisenberg-Euler1936,Marklund2006}.

A central result of the present formulation is the emergence of a
transition probability
\begin{align}
P_{\alpha\rightarrow\alpha^{\prime}} & \propto|\mathcal{M}{}_{\alpha\rightarrow\alpha^{\prime}}|^{2}
\end{align}
between distinct EM coherent states (Eq.\,(\ref{eq: transition probability of coherent states})).
This quantity provides a natural framework for characterizing coherent-field
depletion and backreaction in free propagation of a laser EM field.
In addition, a weak-field estimate based on the Heisenberg\textendash Euler
effective action suggests that coherent-state transition probabilities
may receive nontrivial vacuum-induced corrections even in the absence
of real particle production. The present formulation therefore offers
a framework in which coherent-state transitions can be conceptually
separated from particle-production channels and may provide a way
to distinguish coherent-field transitions from the reduction in the
vacuum-persistence weight that is conventionally attributed to vacuum
decay \citep{Schwinger1951,Nikishov1970,Dunne2005}. This separation
is absent in the conventional vacuum-persistence picture. From this
perspective, the present theory may be regarded as a generalization
of the conventional vacuum-persistence picture, in which the internal
structure of probability loss can be analyzed with a higher level
of resolution.

As a first qualitative application of the formalism, we further examine
the weak coherent-field limit based on the leading-order Heisenberg\textendash Euler
effective action \citep{Heisenberg-Euler1936,Euler_Kockel1935,Weisskopf1936}.
This analysis illustrates how the depletion indicator is related to
field-dependent modifications of the fluctuation spectrum and provides
an initial connection between coherent-state depletion and vacuum-polarization
effects.

The paper is organized as follows. In Sect.~\ref{sec:Gupta-Bleuler}
we review the Gupta\textendash Bleuler condition and coherent-state
representation of laser EM fields. Section \ref{sec:Operator_formalism}
develops the operator-level formalism of coherent-field QED. Section
\ref{sec:Path-integral} derives the corresponding path-integral representation
with coherent-state boundary conditions. In Sect.~\ref{sec:Coherent_EM_transition}
we obtain the effective action and coherent-state transition amplitude.
Section \ref{sec:Discussion} discusses the resulting effective Maxwell
equation, fluctuation-induced currents, coherent-state transition
probabilities, and a weak-field estimate based on the Heisenberg\textendash Euler
effective action. Finally, Sect.~\ref{sec:Conclusion} summarizes
the main conclusions and future directions.

\section{Gupta\textendash Bleuler Condition and Coherent Laser EM Fields\label{sec:Gupta-Bleuler}}

In the following discussion the EM field $\hat{A}_{\mathrm{H}}=\hat{A}_{(+)\mathrm{H}}+\hat{A}_{(-)\mathrm{H}}$
is assumed to satisfy the Gupta\textendash Bleuler condition \citep{Fradkin-Gitman-Shvartsman1991,Gupta1950,Bleuler1950,Itzykson-Zuber1980}.
Hatted symbols denote operators and the subscript ``H'' refers to
objects in the Heisenberg picture. Using the relativistic normalization
of field operators in SI units, suppose that
\begin{alignat}{1}
\hat{A}_{(+)\mathrm{H}}^{\mu}(x) & =\sqrt{\frac{\hbar}{\varepsilon_{0}}}\sum_{\sigma=1,2}\int\frac{d^{3}\boldsymbol{k}}{(2\pi)^{3}\times2k^{0}}\hat{a}_{\mathrm{H}}(\boldsymbol{k},\sigma;t)\epsilon^{\mu}(\boldsymbol{k},\sigma)e^{i\boldsymbol{k}\cdot\boldsymbol{x}}
\end{alignat}
and $\hat{A}_{(-)\mathrm{H}}(x)=[\hat{A}_{(+)\mathrm{H}}(x)]^{\dagger}$
are the positive-frequency and negative-frequency components of $\hat{A}_{\mathrm{H}}(x)$.
The Gupta\textendash Bleuler condition is imposed as 
\begin{align}
\partial_{\mu}\hat{A}_{(+)\mathrm{H}}^{\mu}(x)|\mathrm{Physical}_{\mathrm{H}}\rangle & =0,\label{eq: Gupta-Bleuler}
\end{align}
where $|\mathrm{Physical}_{\mathrm{H}}\rangle$ denotes any physical
state. 

A coherent state \citep{Glauber1963a,Glauber1963b,Sudarshan1963,Sargent1974,Loudon_2000}
is specified by the following eigenvalue equation
\begin{align}
\hat{A}_{(+)\mathrm{H}}^{\mu}(x)|\alpha_{\mathrm{H}}\rangle & =\mathcal{A}_{(+)}^{\mu}(x)|\alpha_{\mathrm{H}}\rangle.
\end{align}
Hence, the expectation value of the EM field in a coherent state is
\begin{align}
\mathcal{A}^{\mu}(x) & =\langle\alpha_{\mathrm{H}}|\hat{A}_{\mathrm{H}}^{\mu}(x)|\alpha_{\mathrm{H}}\rangle\label{eq:coherent_EM_field_01}\\
 & =\mathcal{A}_{(+)}^{\mu}(x)+\mathcal{A}_{(-)}^{\mu}(x)\label{eq:coherene_EM_field_02}
\end{align}
with $\mathcal{A}_{(-)}^{\mu}(x)=[\mathcal{A}_{(+)}^{\mu}(x)]^{*}$.
For an appropriate function $\alpha(\boldsymbol{k},\sigma;t)=\langle\alpha_{\mathrm{H}}|\hat{a}_{\mathrm{H}}(\boldsymbol{k},\sigma;t)|\alpha_{\mathrm{H}}\rangle$,
the coherent state $|\alpha_{\mathrm{H}}\rangle$ defines the set
of $\{\alpha(\boldsymbol{k},\sigma;t)\}$, equivalently the corresponding
spacetime configuration of $\mathcal{A}$ is defined uniquely. For
each mode, the photon-number distribution is not an additional independent
datum once the coherent-state label has been specified. It is fixed
by the coherent state itself and is given by the Poissonian statistics
associated with the corresponding mode amplitude $\alpha(\boldsymbol{k},\sigma;t)$.
Thus, the coherent-state label determines both the coherent-field
expectation value and the photon statistics of the coherent state.
We regard $\mathcal{A}$ as a spatio-temporal laser profile defined
quantum mechanically via the relation between $|\alpha_{\mathrm{H}}\rangle$
and $\{\alpha(\boldsymbol{k},\sigma;t)\}$. We will call this laser
field $\mathcal{A}$ a ``coherent field'' when emphasizing its association
with the coherent state $|\alpha_{\mathrm{H}}\rangle$. By contrast,
the same function may be called a ``prescribed background field''
when it is introduced by hand from the outset. When we adopt the Gupta\textendash Bleuler
condition for a coherent state, i.e., $|\mathrm{Physical}_{\mathrm{H}}\rangle=|\alpha_{\mathrm{H}}\rangle$
in Eq.\,(\ref{eq: Gupta-Bleuler}), the coherent field $\mathcal{A}^{\mu}(x)$
automatically satisfies the Lorenz gauge condition 
\begin{align}
\partial_{\mu}\mathcal{A}^{\mu}(x) & =0.\label{eq: Lorenz condition for a laser pulse}
\end{align}
The coherent-state/coherent-field construction is formulated after
choosing a standard covariant quantization of the EM field (Gupta\textendash Bleuler
quantization). In the following, we restrict the coherent component
to the physical transverse photon sector after imposing the Gupta\textendash Bleuler
condition. In this formulation, the coherent field $\mathcal{A}$
should be understood as a background field defined within the Gupta\textendash Bleuler
quantization scheme. Residual gauge transformations preserving the
Lorenz gauge condition correspond to alternative representatives of
the same gauge orbit. The coherent-state/coherent-field correspondence
established below is formulated within a fixed representative of this
gauge class. The purpose of the present work is not to construct gauge-invariant
background fields, but to establish the correspondence between prescribed
background-field configurations and coherent-state boundary conditions
within a given quantization scheme.

The coherent state can be generated from the vacuum state $|0_{\mathrm{H}}\rangle$
using the displacement operator $\hat{D}_{\mathrm{H}}(\alpha,t)$,
so that 
\begin{align}
|\alpha_{\mathrm{H}}\rangle & =\hat{D}_{\mathrm{H}}(\alpha,t)|0_{\mathrm{H}}\rangle,\label{eq: coherent state by displacement}
\end{align}
\begin{align}
\hat{D}_{\mathrm{H}}(\alpha,t) & =\exp\sum_{\sigma=1,2}\int\frac{d^{3}\boldsymbol{k}}{(2\pi)^{3}\times2k^{0}}[\alpha(\boldsymbol{k},\sigma;t)\hat{a}_{\mathrm{H}}^{\dagger}(\boldsymbol{k},\sigma;t)-\alpha^{*}(\boldsymbol{k},\sigma;t)\hat{a}_{\mathrm{H}}(\boldsymbol{k},\sigma;t)].\label{eq: Displacement operator expansion}
\end{align}
Although the displacement operator is expressed using time-dependent
Heisenberg operators, the coherent state itself remains time independent
in the Heisenberg picture. Accordingly, we suppress the explicit time
label and simply write $\hat{D}_{\mathrm{H}}(\alpha)=\hat{D}_{\mathrm{H}}(\alpha,t)$.
Applying the displacement operator to the EM field operator yields
\begin{align}
\hat{D}_{\mathrm{H}}^{\dagger}(\alpha)\hat{A}_{\mathrm{H}}^{\mu}(x)\hat{D}_{\mathrm{H}}(\alpha) & =\hat{A}_{\mathrm{H}}^{\mu}(x)+\mathcal{A}^{\mu}(x)\hat{\mathbb{I}},\label{eq: Displacement of EM field Heisenberg}
\end{align}
where $\hat{\mathbb{I}}$ is the identity operator \citep{Glauber1963b,Loudon_2000,Frantz1965}.
This relation shows that the coherent field $\mathcal{A}$ appears
as a c-number shift of the quantum EM field operator $\hat{A}_{\mathrm{H}}$.
In the subsequent analysis, this section provides the operator-based
definition of a coherent laser EM field. All displacement transformations
are understood within this fixed quantization scheme.

\section{Operator-level formalism of coherent-field QED systems\label{sec:Operator_formalism}}

For clarity, we confirm the state evolution of the QED system with
a background laser EM field in the Schr\"odinger picture, before
discussing laser-depletion effects induced by the QED vacuum. Field
operators are time independent, while the time evolution is entirely
carried by the quantum states \citep{Itzykson-Zuber1980,Sakurai1967}.
Importantly, this construction is performed at the operator level
and does not require modifying the Hamiltonian or introducing an explicitly
time-dependent external field by hand.

\subsection{State evolution with a laser coherent state\label{sec:Evo_coherent_state}}

Within this picture, a background laser spatio-temporal profile is
encoded into a coherent state $|\alpha_{\mathrm{S}}(t)\rangle$ via
Eqs.\,(\ref{eq:coherent_EM_field_01}-\ref{eq:coherene_EM_field_02})
from the outset, where the subscript ``S'' denotes the Schr\"{o}dinger
picture. Particles other than the EM coherent field, i.e., electrons,
positrons, and incoherent photons, are denoted abstractly by $|\Psi_{\mathrm{S}}(t)\rangle$.
Thus, in the Schr\"{o}dinger picture, the state containing both the
coherent laser field and other particles can be written as 
\begin{align}
|\alpha_{\mathrm{S}}(t),\Psi_{\mathrm{S}}(t)\rangle & =\hat{D}_{\mathrm{S}}(\alpha)|\Psi_{\mathrm{S}}(t)\rangle.\label{eq: coherent state separation}
\end{align}
The usual QED Hamiltonian and its density for the state $|\alpha_{\mathrm{S}}(t),\Psi_{\mathrm{S}}(t)\rangle$
in the Schr\"{o}dinger picture are expressed abstractly as 
\begin{align}
\hat{H}_{\mathrm{S}} & =H_{\mathrm{QED}}[\hat{\psi}_{\mathrm{S}},\hat{\bar{\psi}}_{\mathrm{S}},\hat{A}_{\mathrm{S}}]\\
 & =\int d^{3}\boldsymbol{x}\mathcal{H}_{\mathrm{QED}}(\hat{\psi}_{\mathrm{S}}(\boldsymbol{x}),\hat{\bar{\psi}}_{\mathrm{S}}(\boldsymbol{x}),\hat{A}_{\mathrm{S}}(\boldsymbol{x})),
\end{align}
\begin{align}
\hat{\mathcal{H}}_{\mathrm{S}}(\boldsymbol{x}) & =\mathcal{H}_{\mathrm{QED}}(\hat{\psi}_{\mathrm{S}}(\boldsymbol{x}),\hat{\bar{\psi}}_{\mathrm{S}}(\boldsymbol{x}),\hat{A}_{\mathrm{S}}(\boldsymbol{x})).
\end{align}
Under this preparation, the Schr\"{o}dinger equation governing the
``coherent laser field plus other particles'' state is 
\begin{align}
i\hbar\partial_{t}|\alpha_{\mathrm{S}}(t),\Psi_{\mathrm{S}}(t)\rangle & =\hat{H}_{\mathrm{S}}|\alpha_{\mathrm{S}}(t),\Psi_{\mathrm{S}}(t)\rangle,\label{eq: Schrodinger eq original}
\end{align}
\begin{align}
|\alpha_{\mathrm{S}}(t),\Psi_{\mathrm{S}}(t)\rangle & =\hat{U}(t,t_{0})|\alpha_{\mathrm{S}}(t_{0}),\Psi_{\mathrm{S}}(t_{0})\rangle,
\end{align}
\begin{align}
\hat{U}(t,t_{0}) & =\exp\bigg[-\frac{i}{\hbar}\hat{H}_{\mathrm{S}}(t-t_{0})\bigg].\label{eq: Evolution operator in coherent-field QED}
\end{align}
We refer to the present state-evolution model as ``coherent-field
QED'' emphasizing that coherent fields arise in the following transformation
and the present path integral formalism.

\subsection{State evolution with a laser coherent field\label{sec:Evo_coherent_field}}

We now separate $|\alpha_{\mathrm{S}}(t)\rangle$ and focus on the
time evolution of the quantum state $|\Psi_{\mathrm{S}}(t)\rangle$.
Substituting Eq.\,(\ref{eq: coherent state separation}) into Eq.\,(\ref{eq: Schrodinger eq original}),
the evolution equation becomes 
\begin{align}
i\hbar\partial_{t}|\Psi_{\mathrm{S}}(t)\rangle & =\hat{D}_{\mathrm{S}}^{\dagger}(\alpha)\hat{H}_{\mathrm{S}}\hat{D}_{\mathrm{S}}(\alpha)|\Psi_{\mathrm{S}}(t)\rangle\nonumber \\
 & =H_{\mathrm{QED}}[\hat{\psi}_{\mathrm{S}},\hat{\bar{\psi}}_{\mathrm{S}},\hat{D}_{\mathrm{S}}^{\dagger}(\alpha)\hat{A}_{\mathrm{S}}\hat{D}_{\mathrm{S}}(\alpha)]|\Psi_{\mathrm{S}}(t)\rangle,
\end{align}
where the displacement operator in the Schr\"{o}dinger picture is
\begin{align}
\hat{D}_{\mathrm{S}}(\alpha) & =\exp\sum_{\sigma=1,2}\int\frac{d^{3}\boldsymbol{k}}{(2\pi)^{3}\times2k^{0}}[\alpha(\boldsymbol{k},\sigma;t_{0})\hat{a}_{\mathrm{S}}^{\dagger}(\boldsymbol{k},\sigma;t_{0})-\alpha^{*}(\boldsymbol{k},\sigma;t_{0})\hat{a}_{\mathrm{S}}(\boldsymbol{k},\sigma;t_{0})].
\end{align}
Using the formula (\ref{eq: Displacement of EM field Heisenberg})
in the present picture, 
\begin{align}
\hat{D}_{\mathrm{S}}^{\dagger}(\alpha)\hat{A}_{\mathrm{S}}(t_{0},\boldsymbol{x})\hat{D}_{\mathrm{S}}(\alpha) & =\hat{A}_{\mathrm{S}}(t_{0},\boldsymbol{x})+\mathcal{A}(t_{0},\boldsymbol{x})\hat{\mathbb{I}},\label{eq: Displacement of EM field Schrodinger}
\end{align}
\begin{align}
\mathcal{A}(t_{0},\boldsymbol{x}) & =\langle\alpha_{\mathrm{S}}(t_{0})|\hat{A}_{\mathrm{S}}(t_{0},\boldsymbol{x})|\alpha_{\mathrm{S}}(t_{0})\rangle,
\end{align}
we conclude 
\begin{align}
i\hbar\partial_{t}|\Psi_{\mathrm{S}}(t)\rangle & =H_{\mathrm{QED}}[\hat{\psi}_{\mathrm{S}},\hat{\bar{\psi}}_{\mathrm{S}},\hat{A}_{\mathrm{S}}+\mathcal{A}(t_{0})\hat{\mathbb{I}}]|\Psi_{\mathrm{S}}(t)\rangle.
\end{align}
Throughout the present discussion, the QED Hamiltonian is understood
to be defined with a fixed operator ordering, or equivalently in a
normal-ordered/renormalized convention. Since the displacement operator
is unitary, the transformation $\hat{D}_{\mathrm{S}}^{\dagger}(\alpha)\hat{H}_{\mathrm{S}}\hat{D}_{\mathrm{S}}(\alpha)$
preserves the operator ordering and shifts the EM field operator as
$\hat{A}_{\mathrm{S}}\rightarrow\hat{A}_{\mathrm{S}}+\mathcal{A}\hat{\mathbb{I}}$.
The fermionic interaction terms are transformed by the same unitary
displacement. Thus the coupling of the charged fields to the EM field
operator is rewritten in terms of the shifted field $\hat{A}_{\mathrm{S}}+\mathcal{A}\hat{\mathbb{I}}$,
without introducing an additional assumption for the interaction Hamiltonian.
In addition to the shifted operator terms, this transformation can
generate coherent-state-dependent c-number contributions. These terms
do not affect the operator structure of the quantum fluctuations and
are included in the coherent-field-dependent c-number part of the
theory. Their path-integral counterpart appears in Sec.\,\ref{sec:Path-integral}
through the coherent-field action and the associated effective action.
Here $t_{0}$ is chosen as the time at which the Schr\"{o}dinger
and Heisenberg pictures are matched, $\mathcal{A}(t_{0})$ symbolically
denotes $\mathcal{A}(t_{0},\boldsymbol{x})$ of the fixed field configuration
at the reference time $t_{0}$. Although one would usually write $\hat{O}_{\mathrm{S}}(\boldsymbol{x})=\hat{O}_{\mathrm{S}}(t_{0},\boldsymbol{x})$,
here we retain the explicit notation. Equation (\ref{eq: Displacement of EM field Schrodinger})
is obtained from Eq.\,(\ref{eq: Displacement of EM field Heisenberg})
via the common relation of an observable, $\hat{O}_{\mathrm{S}}=\hat{O}_{\mathrm{H}}(t_{0})$.
Importantly, the background field $\mathcal{A}$ automatically satisfies
the Lorenz gauge condition (\ref{eq: Lorenz condition for a laser pulse})
via the Gupta\textendash Bleuler condition (\ref{eq: Gupta-Bleuler})
\citep{Gupta1950,Bleuler1950} evaluated at the reference time $t_{0}$.
This is consistent with the gauge conditions commonly imposed on prescribed
background fields in strong-field QED, including the construction
of Volkov solutions \citep{Volkov1936} and related background-field
formulations. In the present construction, the coherent-field profile
is encoded in the boundary coherent state rather than introduced as
an explicitly time-dependent external background. The explicit spacetime
dependence of the classical field emerges only after transforming
to the Heisenberg picture. We therefore write 
\begin{align}
|\Psi_{\mathrm{S}}(t)\rangle & =\hat{U}_{\mathcal{A}}(t,t_{0})|\Psi_{\mathrm{S}}(t_{0})\rangle.\label{eq: State in Schrodinger picture}
\end{align}
The associated evolution operator satisfies 
\begin{align}
i\hbar\partial_{t}\hat{U}_{\mathcal{A}}(t,t_{0}) & =H_{\mathrm{QED}}[\hat{\psi}_{\mathrm{S}},\hat{\bar{\psi}}_{\mathrm{S}},\hat{A}_{\mathrm{S}}+\mathcal{A}(t_{0})\hat{\mathbb{I}}]\hat{U}_{\mathcal{A}}(t,t_{0}),\label{eq: time evolution in Schrodinger eq}
\end{align}
with the formal solution 
\begin{align}
\hat{U}_{\mathcal{A}}(t,t_{0}) & =\exp\bigg[-\frac{i}{\hbar}H_{\mathrm{QED}}[\hat{\psi}_{\mathrm{S}},\hat{\bar{\psi}}_{\mathrm{S}},\hat{A}_{\mathrm{S}}+\mathcal{A}(t_{0})\hat{\mathbb{I}}](t-t_{0})\bigg].\label{eq: Evolution operator from Schrodinger picture}
\end{align}
Thus the time evolution of the quantum state $|\Psi_{\mathrm{S}}(t)\rangle$,
with the laser coherent state factored out, incorporates the influence
of the laser coherent field ($\mathcal{A}$) through its explicit
appearance in the evolution operator. To emphasize the dependence
on the prescribed coherent field, we denote the state by
\begin{align}
|\Psi_{\mathrm{S}}(t;\mathcal{A})\rangle & =|\Psi_{\mathrm{S}}(t)\rangle
\end{align}
while keeping in mind that no new state is introduced and the notation
merely makes the background-field dependence explicit. In particular,
the vacuum state associated with a fixed coherent field may be denoted
by

\begin{equation}
|0_{\mathrm{S}}(t;\mathcal{A})\rangle
\end{equation}
which corresponds to the vacuum state of the conventional background-field
formulation of strong-field QED. The coherent field $\mathcal{A}(t_{0},\boldsymbol{x})$
is specified as a boundary condition on the quantum state. Its explicit
time dependence is not fundamental, but emerges upon transforming
to the Heisenberg picture, where the same Hamiltonian generates time-dependent
field operators. 

The distinction between the time-evolution operators $\hat{U}(t,t_{0})$
and $\hat{U}_{\mathcal{A}}(t,t_{0})$ is essential, as it reflects
whether the coherent field is treated as part of the quantum state
or as ``fixed'' coherent-field boundary conditions. General boundary
conditions represent the transition from a coherent field $\mathcal{A}$
to another $\mathcal{A}^{\prime}$ (with $\hat{U}(t,t_{0})$ by Eq.\,(\ref{eq: Evolution operator in coherent-field QED})),
while the fixed case is $\mathcal{A}\rightarrow\mathcal{A}$ (with
$\hat{U}_{\mathcal{A}}(t,t_{0})$ by Eq.\,(\ref{eq: Evolution operator from Schrodinger picture})),
in quantum transitions. 

\subsection{Expressions in the Heisenberg picture}

We discussed the Schr\"{o}dinger picture of coherent-field QED to
clarify the behavior of a coherent state of a laser EM field in the
operator-level formalism. The translation into the Heisenberg picture
is performed by using the time evolution operator $\hat{U}(t,t_{0})$
as usual. The state is given by
\begin{align}
|\alpha_{\mathrm{H}},\Psi_{\mathrm{H}}\rangle & =\hat{U}^{-1}(t,t_{0})|\alpha_{\mathrm{S}}(t),\Psi_{\mathrm{S}}(t)\rangle\\
 & =\hat{D}_{\mathrm{H}}(\alpha)|\Psi_{\mathrm{H}}\rangle
\end{align}
as $|\alpha_{\mathrm{S}}(t_{0}),\Psi_{\mathrm{S}}(t_{0})\rangle=|\alpha_{\mathrm{H}},\Psi_{\mathrm{H}}\rangle$
and $\hat{D}_{\mathrm{H}}(\alpha)=\hat{U}^{-1}(t,t_{0})\hat{D}_{\mathrm{S}}(\alpha)\hat{U}(t,t_{0})$.
At this point it is important to distinguish the reference-time coherent
configuration in the Schr\"{o}dinger picture from the Heisenberg-picture
coherent field. In the Schr\"{o}dinger picture, the coherent-state
boundary condition specifies the coherent-field configuration at the
reference time $t_{0}$, which appears in the displaced Hamiltonian
as $\mathcal{A}(t_{0},\boldsymbol{x})$. After transforming to the
Heisenberg picture, the same coherent-state label defines the spacetime-dependent
expectation value $\mathcal{A}^{\mu}(x)=\langle\alpha_{\mathrm{H}}|\hat{A}_{\mathrm{H}}^{\mu}(x)|\alpha_{\mathrm{H}}\rangle$.
Thus the time dependence of the prescribed laser background is carried
by the Heisenberg operator and by the Heisenberg-picture displacement
operator $\hat{D}_{\mathrm{H}}(\alpha)$, rather than being introduced
as an additional explicitly time-dependent external source. 

The expectation value of an observable $O[\hat{A}_{\mathrm{H}}]$
in $|\alpha_{\mathrm{H}},\Psi_{\mathrm{H}}\rangle$ is
\begin{align}
\langle\alpha_{\mathrm{H}},\Psi_{\mathrm{H}}|O[\hat{A}_{\mathrm{H}}]|\alpha_{\mathrm{H}},\Psi_{\mathrm{H}}\rangle & =\langle\Psi_{\mathrm{H}}|O[\hat{A}_{\mathrm{H}}+\mathcal{A}\hat{\mathbb{I}}]|\Psi_{\mathrm{H}}\rangle.
\end{align}
The right-hand side of the above equation explicitly represents the
coherent-field dependence. The state $|\Psi_{\mathrm{H}}\rangle$
appearing on the right-hand side implicitly depends on the prescribed
coherent field $\mathcal{A}$ through the displaced Hamiltonian $H_{\mathrm{QED}}[\hat{\psi}_{\mathrm{H}},\hat{\bar{\psi}}_{\mathrm{H}},\hat{A}_{\mathrm{H}}+\mathcal{A}\hat{\mathbb{I}}]$.
To make this dependence explicit, we define
\begin{align}
|\Psi_{\mathrm{H}}(\mathcal{A})\rangle & =|\Psi_{\mathrm{H}}\rangle.
\end{align}
In this notation, the symbol $|0_{\mathrm{H}}(\mathcal{A})\rangle=\hat{U}_{\mathcal{A}}^{-1}(t,t_{0})|0_{\mathrm{S}}(t;\mathcal{A})\rangle$
denotes the vacuum state associated with the fixed coherent-field
sector $\mathcal{A}$. The vacuum-persistence amplitude of the conventional
background-field formulation \citep{Furry1951,Schwinger1951,Greiner_Muller_Rafelski_1985,Dittrich-Reuter1985}
is then naturally expressed as

\begin{equation}
\langle0_{(\mathrm{out})\mathrm{H}}(\mathcal{A})|0_{(\mathrm{in})\mathrm{H}}(\mathcal{A})\rangle.
\end{equation}
The present formulation generalizes this structure by replacing the
fixed coherent-field sector $\mathcal{A}$ with transitions between
distinct coherent-field sectors, which will be discussed in the following
section.

\section{Coherent-state boundary conditions in path-integral representation\label{sec:Path-integral}}

We reformulate the operator-level coherent-field QED introduced in
the previous section in the path-integral representation. The essential
point is that the classical background field arises from coherent-state
boundary conditions imposed on the EM field. The construction presented
below is inspired by the standard coherent-state path-integral formalism
developed in quantum optics and many-body theory \citep{Klauder-Skagerstam_1985,Negele-Orland1988,Glauber1963a,Glauber1963b,Sudarshan1963}.
In those approaches, coherent states provide an overcomplete basis
and lead to a functional integral over coherent-state labels. The
present formulation differs in that the coherent states are associated
with the physical EM field of QED and are imposed as asymptotic boundary
conditions on the in- and out-states. Rather than introducing an external
classical background from the outset, the EM coherent field emerges
as the expectation value of the gauge-field operator in the corresponding
coherent state. The resulting path integral therefore describes transitions
between distinct coherent-field sectors and provides a natural starting
point for constructing an effective action that incorporates depletion
and backreaction effects.

\subsection{Coherent-state boundary conditions\label{sec:Coherent-state boundary conditions}}

We consider the transition amplitude 
\begin{equation}
\langle\alpha_{(\mathrm{out})\mathrm{H}}^{\prime},\Psi_{(\mathrm{out})\mathrm{H}}^{\prime}|\alpha_{(\mathrm{in})\mathrm{H}},\Psi_{(\mathrm{in})\mathrm{H}}\rangle,\label{eq: general transition amplitude}
\end{equation}
where $\alpha_{(\mathrm{in})}$ and $\alpha_{(\mathrm{out})}^{\prime}$
denote asymptotic coherent in- and out-states associated with the
laser field, while $\Psi_{(\mathrm{in})}$ and $\Psi_{(\mathrm{out})}^{\prime}$
represent the remaining quantum degrees of freedom. 

Using Eq.\,(\ref{eq: Displacement of EM field Heisenberg}), we separate
the total EM field into 
\begin{align}
\hat{A}_{(\mathrm{tot})\mathrm{H}}^{\mu}(x) & =\hat{A}_{\mathrm{H}}^{\mu}(x)+\mathcal{A}^{\mu}(x)\hat{\mathbb{I}},
\end{align}
where $\mathcal{A}^{\mu}(x)$ is the coherent-field expectation value
determined by the asymptotic coherent states,
\begin{align}
\mathcal{A}_{(\mathrm{in})}^{\mu}(x) & =\langle\alpha_{(\mathrm{in})\mathrm{H}}|\hat{A}_{(\mathrm{in})\mathrm{H}}^{\mu}(x)|\alpha_{(\mathrm{in})\mathrm{H}}\rangle\quad\mathrm{for}\quad t\rightarrow-\infty,\\
\mathcal{A}_{(\mathrm{out})}^{\mu}(x) & =\langle\alpha_{(\mathrm{out})\mathrm{H}}^{\prime}|\hat{A}_{(\mathrm{out})\mathrm{H}}^{\mu}(x)|\alpha_{(\mathrm{out})\mathrm{H}}^{\prime}\rangle\quad\mathrm{for}\quad t\rightarrow+\infty.
\end{align}
In the path-integral formulation developed below, these coherent fields
play the role of boundary conditions. The resulting functional integration
is therefore restricted to coherent-field histories connecting the
asymptotic coherent sectors specified by the in- and out-states.

\subsection{Path-integral construction for free EM coherent fields and photons\label{Path-integral-cohe-gamma}}

Before discussing the full QED system, it is useful to examine how
such coherent-state boundary conditions enter the path integral in
the simpler case of the free EM field. The present construction is
closely related to the standard coherent-state path-integral formalism
developed in quantum optics and many-body theory \citep{Klauder-Skagerstam_1985,Negele-Orland1988}.
In those approaches, coherent states provide an overcomplete basis
and lead to a functional integral over coherent-state variables. The
present formulation should not be regarded as a replacement for these
well-established constructions, but rather as an extension adapted
to coherent-state boundary conditions in QED. In the present framework,
the EM field is decomposed into a coherent-field component $\mathcal{A}$
and the remaining quantum fluctuation (incoherent photons) $A$, and
functional integrations are performed over both histories. If the
quantum fluctuation sector ($A$) is neglected and attention is restricted
to the coherent-field sector alone, the resulting construction reduces
to a functional integral over coherent-field histories and therefore
reproduces the characteristic structure of the standard coherent-state
path-integral formulation. The discussion below shows how the coherent-field
sector and the quantum fluctuation sector naturally appear together
in the present formulation.

The free EM Hamiltonian may be written as
\begin{align}
H_{\mathrm{EM}}[\hat{A}_{\mathrm{H}},\hat{\Pi}_{\mathrm{H}}] & =H_{\mathrm{EM}}[\hat{a}_{\mathrm{H}},\hat{a}_{\mathrm{H}}^{\dagger}]
\end{align}
because the EM field has already been expanded in terms of creation
and annihilation operators in the Gupta\textendash Bleuler formulation.
Here, $\hat{\Pi}_{\mathrm{H}}$ denotes the canonical momentum for
$\hat{A}_{\mathrm{H}}$. For this Hamiltonian, we consider the transition
amplitude
\begin{equation}
\langle\alpha_{(\mathrm{out})\mathrm{H}}^{\prime},\gamma_{(\mathrm{out})\mathrm{H}}^{\prime}|\alpha_{(\mathrm{in})\mathrm{H}},\gamma_{(\mathrm{in})\mathrm{H}}\rangle,
\end{equation}
where $|\gamma_{(\mathrm{in})\mathrm{H}}\rangle$ and $|\gamma_{(\mathrm{out})\mathrm{H}}^{\prime}\rangle$
denote arbitrary photon Fock states. To describe simultaneously the
coherent component of the EM field and its quantum fluctuations, we
introduce displaced Fock states
\begin{align}
|\alpha_{\mathrm{H}},\gamma_{\mathrm{H}}\rangle & =\hat{D}_{\mathrm{H}}(\alpha)|\gamma_{\mathrm{H}}\rangle,
\end{align}
where the displacement operator $\hat{D}_{\mathrm{H}}(\alpha)$ was
defined by Eq.\,(\ref{eq: Displacement operator expansion}). These
states provide a convenient basis because they separate the coherent
sector, represented by $\alpha$, from the remaining quantum fluctuations
encoded in the Fock state $|\gamma_{\mathrm{H}}\rangle$. The identity
operator may then be represented formally as
\begin{align}
\hat{\mathbb{I}} & =\frac{1}{\mathcal{N}}\int\mathcal{D}\mu[\alpha]\sum_{\gamma}|\alpha_{\mathrm{H}},\gamma_{\mathrm{H}}\rangle\langle\alpha_{\mathrm{H}},\gamma_{\mathrm{H}}|,
\end{align}
where $\mathcal{D}\mu[\alpha]$ denotes the coherent-state measure
associated with the relativistically normalized photon modes \citep{Klauder-Skagerstam_1985,Negele-Orland1988,Glauber1963b}
and $\mathcal{N}$ is an overall normalization factor. In a finite-volume
regularization, $\mathcal{D}\mu[\alpha]$ reduces to the product measure
$\prod_{(\boldsymbol{k},\sigma)}d^{2}\alpha(\boldsymbol{k},\sigma)/\pi$,
while in the continuum limit the coherent-state labels are understood
with the same relativistic normalization convention as that used in
the Gupta\textendash Bleuler mode expansion. After dividing the time
interval into infinitesimal segments and inserting this resolution
of unity between neighbouring slices, the transition amplitude becomes
a product of short-time kernels of the form
\begin{align}
\langle\alpha_{(n+1)\mathrm{S}},\gamma_{(n+1)\mathrm{S}} & |e^{-iH_{\mathrm{EM}}[\hat{a}_{\mathrm{S}},\hat{a}_{\mathrm{S}}^{\dagger}]\Delta t/\hbar}|\alpha_{(n)\mathrm{S}},\gamma_{(n)\mathrm{S}}\rangle\nonumber \\
 & =\langle\gamma_{(n+1)\mathrm{S}}|\hat{D}_{\mathrm{S}}^{\dagger}(\alpha_{(n+1)})e^{-iH_{\mathrm{EM}}[\hat{a}_{\mathrm{S}},\hat{a}_{\mathrm{S}}^{\dagger}]\Delta t/\hbar}\hat{D}_{\mathrm{S}}(\alpha_{(n)})|\gamma_{(n)\mathrm{S}}\rangle\nonumber \\
 & =\langle\gamma_{(n+1)\mathrm{S}}|[\hat{D}_{\mathrm{S}}^{\dagger}(\alpha_{(n+1)})\hat{D}_{\mathrm{S}}(\alpha_{(n)})][\hat{D}_{\mathrm{S}}^{\dagger}(\alpha_{(n)})e^{-iH_{\mathrm{EM}}[\hat{a}_{\mathrm{S}},\hat{a}_{\mathrm{S}}^{\dagger}]\Delta t/\hbar}\hat{D}_{\mathrm{S}}(\alpha_{(n)})]|\gamma_{(n)\mathrm{S}}\rangle,
\end{align}

The crucial step is the composition law of displacement operators
\citep{Glauber1963b,Sudarshan1963,Loudon_2000},
\begin{align}
\hat{D}_{\mathrm{S}}^{\dagger}(\bar{\beta})\hat{D}_{\mathrm{S}}(\bar{\alpha}) & =e^{\Phi(\bar{\alpha},\bar{\beta})}\hat{D}_{\mathrm{S}}(\bar{\alpha}-\bar{\beta}),
\end{align}
with
\begin{align}
\Phi(\bar{\alpha},\bar{\beta}) & =\frac{1}{2}\sum_{\sigma=1,2}\int\frac{d^{3}\boldsymbol{k}}{(2\pi)^{3}\times2k^{0}}[\bar{\beta}^{*}(\boldsymbol{k},\sigma)\bar{\alpha}(\boldsymbol{k},\sigma)-\bar{\beta}(\boldsymbol{k},\sigma)\bar{\alpha}^{*}(\boldsymbol{k},\sigma)].
\end{align}
For neighbouring time slices, $\bar{\alpha}=\alpha_{(n)}$ and $\bar{\beta}=\alpha_{(n+1)}$,
so that 
\begin{align}
\hat{D}_{\mathrm{S}}^{\dagger}(\alpha_{(n+1)})\hat{D}_{\mathrm{S}}(\alpha_{(n)}) & =e^{\Phi_{(n)}}\hat{D}_{\mathrm{S}}(-\delta\alpha_{(n)}),
\end{align}
where
\begin{align}
\delta\alpha_{(n)}=\alpha_{(n+1)}-\alpha_{(n)},\quad & \Phi_{(n)}=\Phi(\alpha_{(n)},\alpha_{(n+1)}).
\end{align}
The factor $e^{\Phi_{(n)}}$ generates a nontrivial contribution in
the continuum limit,
\begin{align}
\sum_{n}\Phi_{(n)} & \rightarrow\frac{1}{2}\int_{-\infty}^{\infty}dt\sum_{\sigma=1,2}\int\frac{d^{3}\boldsymbol{k}}{(2\pi)^{3}\times2k^{0}}[\alpha^{*}(\boldsymbol{k},\sigma)\dot{\alpha}(\boldsymbol{k},\sigma)-\dot{\alpha}^{*}(\boldsymbol{k},\sigma)\alpha(\boldsymbol{k},\sigma)].
\end{align}

The Hamiltonian contribution is obtained from the second factor in
the kernel decomposition above. Using the displacement transformation,
\begin{align}
\hat{D}_{\mathrm{S}}^{\dagger}(\alpha)H_{\mathrm{EM}}[\hat{a}_{\mathrm{S}},\hat{a}_{\mathrm{S}}^{\dagger}]\hat{D}_{\mathrm{S}}(\alpha) & =H_{\mathrm{EM}}[\hat{a}_{\mathrm{S}}+\alpha\hat{\mathbb{I}},\hat{a}_{\mathrm{S}}^{\dagger}+\alpha^{*}\hat{\mathbb{I}}].
\end{align}
It follows that the coherent variables and the quantum fluctuation
operators enter only through the combination $\hat{a}_{\mathrm{S}}+\alpha\hat{\mathbb{I}}$
and $\hat{a}_{\mathrm{S}}^{\dagger}+\alpha^{*}\hat{\mathbb{I}}$.
At this stage the origin of the two sectors appearing in the present
formulation becomes clear. The functional integration over $\alpha$
and $\alpha^{*}$ originates from the coherent-state degrees of freedom
and describes transitions between different coherent configurations.
By contrast, the summation over photon Fock states represents the
ordinary quantum fluctuations of the EM field. 

In the next step, the coherent-state labels $\alpha$ will be translated
into EM coherent fields through their expectation values, while the
Fock-sector contribution will be converted into the usual field-theoretic
path integral over quantum EM fluctuations. The coherent-state variables
introduced above are not independent physical fields but labels specifying
coherent states of the EM field. Nevertheless, because the EM field
operator has been expanded in terms of creation and annihilation operators,
every coherent-state configuration uniquely determines a corresponding
EM coherent field through its expectation value. We defined the coherent
field $\mathcal{A}$ by Eqs.\,(\ref{eq:coherent_EM_field_01}-\ref{eq:coherene_EM_field_02}),
the positive-frequency component is therefore
\begin{alignat}{1}
\mathcal{A}_{(+)}^{\mu}(x) & =\sqrt{\frac{\hbar}{\varepsilon_{0}}}\sum_{\sigma=1,2}\int\frac{d^{3}\boldsymbol{k}}{(2\pi)^{3}\times2k^{0}}\alpha(\boldsymbol{k},\sigma;t)\epsilon^{\mu}(\boldsymbol{k},\sigma)e^{i\boldsymbol{k}\cdot\boldsymbol{x}},
\end{alignat}
while the negative-frequency component is obtained by complex conjugation.
This establishes a one-to-one correspondence between the coherent-state
history $\alpha(t)$ and the coherent-field history $\mathcal{A}(t)$.
The functional integration over coherent-state labels may therefore
be reinterpreted as a functional integral over coherent-field histories,
\begin{align}
\int\mathcal{D}\mu[\alpha]\quad & \Rightarrow\quad\int\mathcal{D}\mathcal{A}.
\end{align}
The asymptotic coherent states introduced in Sect.\,\ref{sec:Coherent-state boundary conditions}
specify the endpoint coherent-field wave functionals. In the semiclassical
notation used below, this is represented by coherent-field histories
centered on the asymptotic coherent-field configurations,
\begin{align}
\mathcal{A}^{\mu}(x)\rightarrow\mathcal{A}_{(\mathrm{in})}^{\mu}(x),\quad & t\rightarrow-\infty,
\end{align}
\begin{align}
\mathcal{A}^{\mu}(x)\rightarrow\mathcal{A}_{(\mathrm{out})}^{\mu}(x),\quad & t\rightarrow+\infty.
\end{align}
The coherent-sector contribution is then represented symbolically
as
\begin{align}
\int\mathcal{D}\mu[\alpha]\langle\alpha_{(\mathrm{out})\mathrm{H}}^{\prime}|\alpha_{(N)\mathrm{H}}\rangle\langle\alpha_{(0)\mathrm{H}}|\alpha_{(\mathrm{in})\mathrm{H}}\rangle\quad & \Rightarrow\quad\int_{\mathcal{A}_{(\mathrm{in})}}^{\mathcal{A}_{(\mathrm{out})}}\mathcal{D}\mathcal{A}.
\end{align}
We refer to this compact representation as the coherent-field path
integral, because the coherent-state labels are translated into histories
of the corresponding EM coherent field. Here $|\alpha_{(0)\mathrm{H}}\rangle$
and $|\alpha_{(N)\mathrm{H}}\rangle$ denote the coherent states appearing
in the first and last inserted resolutions of identity in the time-sliced
construction. They should not be confused with the prescribed asymptotic
coherent states $|\alpha_{(\mathrm{in})\mathrm{H}}\rangle$ and $|\alpha_{(\mathrm{out})\mathrm{H}}^{\prime}\rangle$
themselves. Strictly speaking, the endpoint overlap factors $\langle\alpha_{(0)\mathrm{H}}|\alpha_{(\mathrm{in})\mathrm{H}}\rangle$
and $\langle\alpha_{(\mathrm{out})\mathrm{H}}^{\prime}|\alpha_{(N)\mathrm{H}}\rangle$
do not impose exact boundary values of the coherent field, since coherent
states are not mutually orthogonal. Rather, they define Gaussian endpoint
weights centered around the asymptotic coherent fields $\mathcal{A}_{(\mathrm{in})}$
and $\mathcal{A}_{(\mathrm{out})}$. More explicitly, the corresponding
initial endpoint weight $|\langle\alpha_{(0)\mathrm{H}}|\alpha_{(\mathrm{in})\mathrm{H}}\rangle|^{2}$
is
\begin{align}
|\langle\alpha_{(0)\mathrm{H}}|\alpha_{(\mathrm{in})\mathrm{H}}\rangle|^{2} & =\exp\bigg[-\sum_{\sigma=1,2}\int\frac{d^{3}\boldsymbol{k}}{(2\pi)^{3}\times2k^{0}}|\alpha_{(0)}(\boldsymbol{k},\sigma)-\alpha_{(\mathrm{in})}(\boldsymbol{k},\sigma)|^{2}\bigg].
\end{align}
Here $\alpha_{(0)}(\boldsymbol{k},\sigma)$ and $\alpha_{(\mathrm{in})}(\boldsymbol{k},\sigma)$
are the Fourier-mode coefficients, equivalently the coherent-state
labels, of the positive-frequency coherent fields $\mathcal{A}_{(0)(+)}^{\mu}(x)$
and $\mathcal{A}_{(\mathrm{in})(+)}^{\mu}(x)$ in the relativistically
normalized expansion 
\begin{alignat}{1}
\mathcal{A}_{(j)(+)}^{\mu}(x) & =\sqrt{\frac{\hbar}{\varepsilon_{0}}}\sum_{\sigma=1,2}\int\frac{d^{3}\boldsymbol{k}}{(2\pi)^{3}\times2k^{0}}\alpha_{(j)}(\boldsymbol{k},\sigma)\epsilon^{\mu}(\boldsymbol{k},\sigma)e^{-ik_{\nu}x^{\nu}},\quad j=0,\,\mathrm{in}.
\end{alignat}
The initial endpoint factor $|\langle\alpha_{(0)\mathrm{H}}|\alpha_{(\mathrm{in})\mathrm{H}}\rangle|^{2}$
is maximized when $\alpha_{(0)}(\boldsymbol{k},\sigma)=\alpha_{(\mathrm{in})}(\boldsymbol{k},\sigma)$,
and it exponentially suppresses configurations for which the difference
$\alpha_{(0)}-\alpha_{(\mathrm{in})}$ carries a large photon number.
Through the one-to-one correspondence between these Fourier-mode coefficients
and the positive-frequency coherent field, this endpoint factor selects
coherent-field histories centered around the prescribed asymptotic
field $\mathcal{A}_{(\mathrm{in})}$. Similarly, the final endpoint
factor $|\langle\alpha_{(\mathrm{out})\mathrm{H}}^{\prime}|\alpha_{(N)\mathrm{H}}\rangle|^{2}$
is a Gaussian weight centered around $\alpha_{(N)}(\boldsymbol{k},\sigma)=\alpha_{(\mathrm{out})}^{\prime}(\boldsymbol{k},\sigma)$,
and it suppresses coherent-field histories whose final coherent configuration
differs appreciably from the prescribed asymptotic field $\mathcal{A}_{(\mathrm{out})}$.
In this sense, we use the compact notation $\int_{\mathcal{A}_{(\mathrm{in})}}^{\mathcal{A}_{(\mathrm{out})}}\mathcal{D}\mathcal{A}$
as a shorthand for a functional integral over coherent-field histories
weighted by these endpoint wave functionals. The asymptotic coherent
states play the role of boundary wave functionals selecting the coherent-field
history in this representation. They are not introduced as an additional
external source term; rather, the coherent-field configuration is
selected by the initial and final coherent-state overlap factors.
The path integral is still rooted in the standard coherent-state path-integral
construction, but its integration variable is reinterpreted as a coherent-field
history rather than as an abstract coherent-state label. The remaining
factor in the path integral originates from the photon Fock states
$|\gamma\rangle$. Since these states span the ordinary EM Hilbert
space, the corresponding summation generates the standard field-theoretic
path integral describing quantum EM fluctuations. At the operator
level, the coherent-state variables and the fluctuation operators
appear only through the displaced combination $\hat{a}_{\mathrm{S}}+\alpha\hat{\mathbb{I}}$
and $\hat{a}_{\mathrm{S}}^{\dagger}+\alpha^{*}\hat{\mathbb{I}}$.
Passing from the mode representation to the field representation,
this combination corresponds to $\hat{A}_{\mathrm{S}}+\mathcal{A}\hat{\mathbb{I}}$
reduced to c-number fields in the path integral. The total EM field
therefore decomposes as
\begin{align}
A_{\mathrm{Total}}^{\mu}(x) & =A^{\mu}(x)+\mathcal{A}^{\mu}(x),
\end{align}
where $A$ represents the quantum fluctuation field and $\mathcal{A}$
represents the coherent-field component associated with the coherent-field
history. The decomposition $A_{\mathrm{Total}}=A+\mathcal{A}$ is
not introduced as an arbitrary split of a c-number field configuration.
It is defined by the displaced-state basis: $\mathcal{A}$ labels
the coherent-state sector, whereas $A$ represents fluctuations in
the displaced Fock sector with vanishing coherent expectation value.

The factors $e^{\Phi_{(n)}}$ and $\hat{D}_{\mathrm{S}}(-\delta\alpha_{(n)})$
play an essential role in reconstructing the phase-space structure
associated with the coherent field. The displaced Hamiltonian contains
the instantaneous dependence on the coherent-field configuration $\mathcal{A}_{(n)}$,
but the terms associated with the change of the coherent-state label
between neighbouring time slices do not arise from the Hamiltonian
alone. They are supplied by the overlap between neighbouring displaced
Fock-state bases. In particular, $e^{\Phi_{(n)}}$ gives the c-number
part of the phase-space term for the coherent component, while $\hat{D}_{\mathrm{S}}(-\delta\alpha_{(n)})$
contributes the operator-valued part required when the displaced fluctuation
basis is changed from one time slice to the next. Together with the
short-time evolution generated by the displaced Hamiltonian, these
terms complete the exponent of the time-sliced kernel into the phase-space
action of the EM field. Writing $\mathcal{A}(t_{(n)},\boldsymbol{x})=\mathcal{A}_{(n)}(\boldsymbol{x})$,
the sum of the time-sliced exponents contains the coherent-field phase-space
contribution $\sum_{n}\Delta t\int d^{3}\boldsymbol{x}\Pi_{\mathcal{A}}(t_{(n)},\boldsymbol{x})\dot{\mathcal{A}}(t_{(n)},\boldsymbol{x})$,
together with the Hamiltonian evaluated on the displaced field variables.
Equivalently, in the continuum limit the phase-space action can be
written in terms of the total EM field, $A_{\mathrm{Total}}$, and
its conjugate momentum. After the canonical momentum is eliminated
in the standard way, this expression becomes the free EM action evaluated
on the total field, $S_{\mathrm{EM}}[A_{\mathrm{Total}}]=S_{\mathrm{EM}}[A+\mathcal{A}]$.
The transition amplitude of the free EM field acquires the schematic
form,
\begin{align}
\langle\alpha_{(\mathrm{out})\mathrm{H}}^{\prime},\gamma_{(\mathrm{out})\mathrm{H}}^{\prime}|\alpha_{(\mathrm{in})\mathrm{H}},\gamma_{(\mathrm{in})\mathrm{H}}\rangle & =\int_{\mathcal{A}_{(\mathrm{in})}}^{\mathcal{A}_{(\mathrm{out})}}\mathcal{D}\mathcal{A}\int\mathcal{D}A\gamma_{(\mathrm{out})}^{\prime*}[A_{(\mathrm{out})}]\gamma_{(\mathrm{in})}[A_{(\mathrm{in})}]\exp\frac{i}{\hbar c}S_{\mathrm{EM}}[A+\mathcal{A}],
\end{align}
\begin{align}
\gamma_{(\mathrm{in})}[A_{(0)}] & =\langle A_{(0)\mathrm{H}}|\gamma_{(\mathrm{in})\mathrm{H}}\rangle,
\end{align}
\begin{align}
\gamma_{(\mathrm{out})}^{\prime*}[A_{(N)}] & =\langle\gamma_{(\mathrm{out})\mathrm{H}}^{\prime}|A_{(N)\mathrm{H}}\rangle.
\end{align}
Here we use the gauge-fixed free EM action in the Feynman gauge \citep{Fradkin-Gitman-Shvartsman1991,Itzykson-Zuber1980,Nakanishi-Ojima1990},
\begin{align}
S_{\mathrm{EM}}[A+\mathcal{A}] & =-\int d^{4}x\frac{1}{2\mu_{0}}\partial_{\mu}[A_{\nu}(x)+\mathcal{A}_{\nu}(x)]\partial^{\mu}[A^{\nu}(x)+\mathcal{A}^{\nu}(x)].
\end{align}
This form is equivalent, up to a surface term, to the Maxwell action
supplemented by the Lorenz-gauge-fixing term with gauge parameter
$\xi=1$; see also Eq.\,(\ref{eq: QED Lagrangian density}). This
expression already exhibits the essential structure of the present
formulation. The functional integration over $A$ describes the ordinary
quantum fluctuations of the EM field. The functionals $\gamma_{(\mathrm{in})}[A_{(0)}]$
and $\gamma_{(\mathrm{out})}^{\prime*}[A_{(N)}]$ identify the boundary
conditions for $A$. By contrast, the functional integration over
$\mathcal{A}$ originates from coherent-state boundary conditions
and parametrizes transitions between distinct coherent-field sectors.
The latter should not be interpreted as an additional integration
over microscopic photon degrees of freedom, which are already accounted
for by the integration over $A$. Rather, it represents a summation
over possible coherent-field histories connecting the asymptotic coherent
states. 

The free-field model therefore reveals the fundamental structure of
coherent-field QED as a double functional integral: one functional
integration describes quantum fluctuations around a given coherent-field
history, while the other describes transitions between different coherent-field
sectors specified by coherent-state boundary conditions.

\subsection{Path-integral representation of coherent-field QED}

The construction developed in the previous subsection can be extended
straightforwardly to the full QED system. Instead of restricting the
fluctuation sector to photon Fock states, we now consider arbitrary
QED states containing photons, electrons, positrons, and their multiparticle
excitations.

The transition amplitude introduced in Sect.\,\ref{sec:Coherent-state boundary conditions}
is 
\[
\langle\alpha_{(\mathrm{out})\mathrm{H}}^{\prime},\Psi_{(\mathrm{out})\mathrm{H}}^{\prime}|\alpha_{(\mathrm{in})\mathrm{H}},\Psi_{(\mathrm{in})\mathrm{H}}\rangle,\tag{{\ref{eq: general transition amplitude}}}
\]
where the coherent states specify the asymptotic EM coherent field,
while the states $|\Psi\rangle$ describe the remaining QED degrees
of freedom. As in the free-field construction, we insert complete
sets of displaced QED states,
\begin{align}
|\alpha_{\mathrm{H}},\Psi_{\mathrm{H}}\rangle & =\hat{D}_{\mathrm{H}}(\alpha)|\Psi_{\mathrm{H}}\rangle,
\end{align}
\begin{align}
\hat{\mathbb{I}} & =\frac{1}{\mathcal{N}}\int\mathcal{D}\mu[\alpha]\sum_{\Psi}|\alpha_{\mathrm{H}},\Psi_{\mathrm{H}}\rangle\langle\alpha_{\mathrm{H}},\Psi_{\mathrm{H}}|.
\end{align}
The coherent-state sector proceeds exactly as in Sect.\,\ref{Path-integral-cohe-gamma}.
The overlap of neighbouring coherent states generates the factors
$e^{\Phi_{(n)}}$ and $\hat{D}_{\mathrm{S}}(-\delta\alpha_{(n)})$,
while the QED Hamiltonian is displaced according to 
\begin{align}
\hat{D}_{\mathrm{S}}^{\dagger}(\alpha)H_{\mathrm{QED}}[\hat{\psi}_{\mathrm{S}},\hat{\bar{\psi}}_{\mathrm{S}},\hat{A}_{\mathrm{S}},\hat{\Pi}_{\mathrm{S}}]\hat{D}_{\mathrm{S}}(\alpha) & =H_{\mathrm{QED}}[\hat{\psi}_{\mathrm{S}},\hat{\bar{\psi}}_{\mathrm{S}},\hat{A}_{\mathrm{S}}+\mathcal{A}\hat{\mathbb{I}},\hat{\Pi}_{\mathrm{S}}+\Pi_{\mathcal{A}}\hat{\mathbb{I}}]
\end{align}
since the displacement operator acts only on the EM field degrees
of freedom. After keeping the endpoint overlap factors explicitly,
the coherent-state labels may again be reinterpreted as coherent-field
histories,
\begin{align}
\int\mathcal{D}\mu[\alpha]\langle\alpha_{(\mathrm{out})\mathrm{H}}^{\prime}|\alpha_{(N)\mathrm{H}}\rangle\langle\alpha_{(0)\mathrm{H}}|\alpha_{(\mathrm{in})\mathrm{H}}\rangle\quad & \Rightarrow\quad\int_{\mathcal{A}_{(\mathrm{in})}}^{\mathcal{A}_{(\mathrm{out})}}\mathcal{D}\mathcal{A}.
\end{align}
This $\mathcal{A}$ integration is the coherent-field path-integral
sector of the full QED transition amplitude, with its endpoint weights
specified by the asymptotic coherent-state overlap factors. The remaining
QED sector generates the ordinary field-theoretic path integral, 
\begin{align}
\sum_{\Psi}\quad & \Rightarrow\quad\int\mathcal{D}A\mathcal{D}\psi\mathcal{D}\bar{\psi}.
\end{align}
The total EM field decomposes into
\begin{align}
A_{\mathrm{Total}}^{\mu}(x) & =A^{\mu}(x)+\mathcal{A}^{\mu}(x),
\end{align}
where $A$ denotes the quantum fluctuation field and $\mathcal{A}$
denotes the coherent-field component associated with the coherent-field
history.

Proceeding as in the ordinary construction of the QED path integral
\citep{Fradkin-Gitman-Shvartsman1991,Itzykson-Zuber1980,Dittrich-Reuter1985},
the continuum limit yields
\begin{align}
\langle\alpha_{(\mathrm{out})\mathrm{H}}^{\prime}, & \Psi_{(\mathrm{out})\mathrm{H}}^{\prime}|\alpha_{(\mathrm{in})\mathrm{H}},\Psi_{(\mathrm{in})\mathrm{H}}\rangle\nonumber \\
 & =\int_{\mathcal{A}_{(\mathrm{in})}}^{\mathcal{A}_{(\mathrm{out})}}\mathcal{D}\mathcal{A}\int\mathcal{D}A\mathcal{D}\psi\mathcal{D}\bar{\psi}\Psi_{(\mathrm{out})}^{\prime*}[A_{(N)},\psi_{(N)},\bar{\psi}_{(N)}]\Psi_{(\mathrm{in})}[A_{(0)},\psi_{(0)},\bar{\psi}_{(0)}]\nonumber \\
 & \quad\exp\bigg[\frac{i}{\hbar c}\int d^{4}x\mathcal{L}_{\mathrm{QED}}(\psi(x),\bar{\psi}(x),A(x)+\mathcal{A}(x))\bigg].\label{eq: general scattering amplitude}
\end{align}
\begin{align}
\Psi_{(\mathrm{in})}[A_{(0)},\psi_{(0)},\bar{\psi}_{(0)}] & =\langle A_{(0)\mathrm{H}},\psi_{(0)\mathrm{H}},\bar{\psi}_{(0)\mathrm{H}}|\Psi_{(\mathrm{in})\mathrm{H}}\rangle,
\end{align}
\begin{align}
\Psi_{(\mathrm{out})}^{\prime*}[A_{(N)},\psi_{(N)},\bar{\psi}_{(N)}] & =\langle\Psi_{(\mathrm{out})\mathrm{H}}^{\prime}|A_{(N)\mathrm{H}},\psi_{(N)\mathrm{H}},\bar{\psi}_{(N)\mathrm{H}}\rangle.
\end{align}
Suppose the QED Lagrangian density under the Feynman gauge ($\xi=1$)
\citep{Fradkin-Gitman-Shvartsman1991,Itzykson-Zuber1980} is 
\begin{align}
\mathcal{L}_{\mathrm{QED}}(\psi,\bar{\psi},A) & =c\bar{\psi}\gamma^{\mu}[i\hbar\partial_{\mu}+eA_{\mu}]\psi-mc^{2}\bar{\psi}\psi-\frac{1}{4\mu_{0}}F_{\mu\nu}F^{\mu\nu}-\frac{1}{2\mu_{0}\xi}(\partial_{\mu}A^{\mu})^{2},\label{eq: QED Lagrangian density}
\end{align}
where $e>0$ denotes the magnitude of the elementary charge; the electron
charge is $-e$. This expression is the fundamental coherent-field
path-integral representation of coherent-field QED. The functional
integration over $A$, $\psi$, and $\bar{\psi}$ describes the ordinary
quantum fluctuations of QED, while the additional functional integration
over $\mathcal{A}$ originates from coherent-state boundary conditions
and parametrizes transitions between distinct coherent-field sectors.
Unlike the conventional background-field formulation, the EM coherent
field is therefore not introduced as a prescribed external field.
Instead, it is treated as a dynamical history over which the transition
amplitude is evaluated. The path integral above provides the starting
point for constructing an effective action that simultaneously incorporates
the quantum dynamics of QED fluctuations and the evolution of the
EM coherent field, including depletion and backreaction effects.

Consider now the special case in which the asymptotic coherent-state
labels coincide, $\alpha^{\prime}=\alpha$. In this case, the corresponding
coherent-state boundary conditions imply identical asymptotic coherent-field
expectation values, 
\begin{align}
\mathcal{A}_{(\mathrm{in})}^{\mu}(x)=\bar{\mathcal{A}}^{\mu}(x),\quad & t\rightarrow-\infty,
\end{align}
\begin{align}
\mathcal{A}_{(\mathrm{out})}^{\mu}(x)=\bar{\mathcal{A}}^{\mu}(x),\quad & t\rightarrow+\infty.
\end{align}
However, this condition constrains only the asymptotic coherent-state
expectation values and does not uniquely determine the intermediate
coherent-field history. Therefore, the functional integration over
$\mathcal{A}$ remains present even when $\alpha^{\prime}=\alpha$.

To recover the conventional background-field formulation, one must
impose a stronger condition. Specifically, one must treat a single
coherent-field history $\bar{\mathcal{A}}$ rather than summing over
all coherent-field histories consistent with the asymptotic boundary
conditions. In this limit, the coherent-field history is fixed and
the functional integration over $\mathcal{A}$ contributes only an
overall normalization factor. The transition amplitude then reduces,
up to an overall normalization factor, to the vacuum-persistence amplitude
\begin{align}
\langle0_{(\mathrm{out})\mathrm{H}}(\bar{\mathcal{A}})|0_{(\mathrm{in})\mathrm{H}}(\bar{\mathcal{A}})\rangle & =\int_{(0_{(\mathrm{in})}^{A},0_{(\mathrm{in})}^{\psi,\bar{\psi}})}^{(0_{(\mathrm{out})}^{A},0_{(\mathrm{out})}^{\psi,\bar{\psi}})}\mathcal{D}A\mathcal{D}\psi\mathcal{D}\bar{\psi}\exp\bigg[\frac{i}{\hbar c}\int d^{4}x\mathcal{L}_{\mathrm{QED}}(\psi(x),\bar{\psi}(x),A(x)+\bar{\mathcal{A}}(x))\bigg].
\end{align}
This expression coincides with the vacuum-persistence amplitude of
the conventional background-field formulation of QED, in which $\bar{\mathcal{A}}$
is regarded as a prescribed external field \citep{Furry1951,Ritus1985,Piazza2012,Fedotov2023}.
It is precisely the generating functional underlying the Furry-picture
formulation and the derivation of the Heisenberg\textendash Euler
effective action \citep{Schwinger1951,Greiner_Muller_Rafelski_1985,Furry1951,Dittrich-Reuter1985}.
It should therefore be emphasized that the condition $\alpha^{\prime}=\alpha$
by itself does not recover the conventional background-field description.
The latter is obtained only after imposing the additional restriction
that a single coherent-field history $\bar{\mathcal{A}}$ is prescribed,
rather than being integrated over in the path integral.

The present formalism therefore extends the conventional background-field
framework by allowing transitions between distinct coherent-field
sectors, $\alpha\rightarrow\alpha^{\prime}$ in the state language,
for which the coherent-field history is not prescribed a priori. Such
transitions naturally provide a framework for describing depletion,
backreaction, and the dynamical evolution of the EM coherent field
itself \citep{Ilderton-Seipt2018,Adamo-Ilderton,Fedotov2023}.

\section{Coherent-field transition in the QED vacuum\label{sec:Coherent_EM_transition}}

We consider the following amplitude in the present theoretical framework:
\begin{align}
\langle\alpha_{(\mathrm{out})\mathrm{H}}^{\prime},0_{(\mathrm{out})\mathrm{H}}^{A,\psi,\bar{\psi}}|\alpha_{(\mathrm{in})\mathrm{H}},0_{(\mathrm{in})\mathrm{H}}^{A,\psi,\bar{\psi}}\rangle & =\int_{\mathcal{A}_{(\mathrm{in})}}^{\mathcal{A}_{(\mathrm{out})}}\mathcal{D}\mathcal{A}\int_{(0_{(\mathrm{in})}^{A},0_{(\mathrm{in})}^{\psi,\bar{\psi}})}^{(0_{(\mathrm{out})}^{A},0_{(\mathrm{out})}^{\psi,\bar{\psi}})}\mathcal{D}A\mathcal{D}\psi\mathcal{D}\bar{\psi}\nonumber \\
 & \quad\exp\bigg[\frac{i}{\hbar c}\int d^{4}x\mathcal{L}_{\mathrm{QED}}(\psi(x),\bar{\psi}(x),A(x)+\mathcal{A}(x))\bigg],\label{eq: coherent state change in vacuum 01}
\end{align}
This amplitude represents the vacuum persistence amplitude between
two asymptotic coherent-field sectors. In contrast to the conventional
background-field formulations of QED, where the background field is
fixed externally from the outset, the present formulation allows transitions
between distinct coherent laser configurations through the functional
integration over $\mathcal{A}$. The coherent field therefore remains
connected to the underlying quantum description of the EM field via
asymptotic coherent-state boundary conditions.

This amplitude illustrates the transition from a laser EM coherent
state $\alpha$ to another state $\alpha^{\prime}$. We define $D_{\mu}(x)=\partial_{\mu}-i(e/\hbar)[A_{\mu}(x)+\mathcal{A}_{\mu}(x)]$
for $\mathcal{L}_{\mathrm{QED}}(\psi(x),\bar{\psi}(x),A(x)+\mathcal{A}(x))$
in Eq.\,(\ref{eq: coherent state change in vacuum 01}). Under the
Feynman gauge, the Dirac-field contribution is Gaussian in the Grassmann
variables and can therefore be integrated out formally. Here the vacuum
boundary conditions $0_{(\mathrm{in})}^{\psi,\bar{\psi}}$ and $0_{(\mathrm{out})}^{\psi,\bar{\psi}}$
denote the asymptotic fermionic vacuum sectors associated with the
fluctuation fields. The corresponding path integral becomes
\begin{align}
\int_{0_{(\mathrm{in})}^{\psi,\bar{\psi}}}^{0_{(\mathrm{out})}^{\psi,\bar{\psi}}}\mathcal{D}\psi\mathcal{D}\bar{\psi}\exp\frac{i}{\hbar c}\int d^{4}x\bar{\psi}(x)(i\hbar c\gamma^{\mu}D_{\mu}(x)-mc^{2}\mathbb{I})\psi(x) & =N\det\bigg(i\gamma^{\mu}D_{\mu}-\frac{mc}{\hbar}\mathbb{I}\bigg)\nonumber \\
 & =N\exp\frac{i}{\hbar c}\Gamma_{\psi\bar{\psi}}[A+\mathcal{A}],
\end{align}
where $N$ is a field-independent normalization constant. The resulting
functional determinant defines the one-loop effective action $\Gamma_{\psi\bar{\psi}}[A+\mathcal{A}]$
which describes vacuum polarization \citep{Schwinger1951,Greiner_Muller_Rafelski_1985,Dittrich-Gies2000,Ritus1985,Dunne2005,Klauder-Skagerstam_1985,Negele-Orland1988,Dittrich-Reuter1985},
which depends on the total gauge field configuration $A+\mathcal{A}$.
Substituting the fermionic functional determinant into the full transition
amplitude yields an effective description involving only the gauge-field
degrees of freedom:
\begin{align}
\langle\alpha_{(\mathrm{out})\mathrm{H}}^{\prime},0_{(\mathrm{out})\mathrm{H}}^{A,\psi,\bar{\psi}}|\alpha_{(\mathrm{in})\mathrm{H}},0_{(\mathrm{in})\mathrm{H}}^{A,\psi,\bar{\psi}}\rangle & =N\int_{\mathcal{A}_{(\mathrm{in})}}^{\mathcal{A}_{(\mathrm{out})}}\mathcal{D}\mathcal{A}\int_{0_{(\mathrm{in})}^{A}}^{0_{(\mathrm{out})}^{A}}\mathcal{D}A\exp\frac{i}{\hbar c}\bigg\{ S_{\mathrm{EM}}[A+\mathcal{A}]+\Gamma_{\psi\bar{\psi}}[A+\mathcal{A}]\bigg\}.\label{eq: coherent state change in vacuum 02}
\end{align}
The remaining functional integrations retain two distinct structures:
the integration over the coherent field $\mathcal{A}$ interpolates
between asymptotic coherent-field sectors corresponding to the laser
field, while the integration over $A$ represents the quantum fluctuation
of the EM field around those coherent configurations. The resulting
expression is therefore a coherent-field generalization of the usual
vacuum persistence amplitude underlying the Heisenberg\textendash Euler
effective action. In the special case where the quantum-photon fluctuation
field $A$ is neglected, the coherent field is fixed ($\mathcal{A}_{(\mathrm{out})}=\mathcal{A}_{(\mathrm{in})}$)
and the integration over $\mathcal{A}$ is suppressed, the present
formulation reduces to the conventional background-field treatment
of strong-field QED and reproduces the Heisenberg\textendash Euler
effective action. In contrast, allowing transitions
\begin{align}
\mathcal{A}_{(\mathrm{in})} & \rightarrow\mathcal{A}_{(\mathrm{out})}
\end{align}
naturally incorporates depletion and backreaction effects at the level
of asymptotic coherent states.

We expand $S_{\mathrm{EM}}[A+\mathcal{A}]+\Gamma_{\psi\bar{\psi}}[A+\mathcal{A}]$
around the coherent field $\mathcal{A}$:
\begin{align}
\langle\alpha_{(\mathrm{out})\mathrm{H}}^{\prime},0_{(\mathrm{out})\mathrm{H}}^{A,\psi,\bar{\psi}}|\alpha_{(\mathrm{in})\mathrm{H}},0_{(\mathrm{in})\mathrm{H}}^{A,\psi,\bar{\psi}}\rangle & =N\int_{\mathcal{A}_{(\mathrm{in})}}^{\mathcal{A}_{(\mathrm{out})}}\mathcal{D}\mathcal{A}\exp\frac{i}{\hbar c}\{S_{\mathrm{EM}}[\mathcal{A}]+\Gamma_{\psi\bar{\psi}}[\mathcal{A}]+\Gamma_{A}[\mathcal{A}]\},\label{eq: coherent state change in vacuum 03}
\end{align}
\begin{align}
\Gamma_{A}[\mathcal{A}] & =-i\hbar c\ln\int\mathcal{D}A\exp\frac{i}{\hbar c}\bigg[\int d^{4}x\frac{\delta\{S_{\mathrm{EM}}[\mathcal{A}]+\Gamma_{\psi\bar{\psi}}[\mathcal{A}]\}}{\delta\mathcal{A}^{\mu}(x)}A^{\mu}(x)\nonumber \\
 & \quad+\frac{1}{2}\int d^{4}x\int d^{4}y\frac{\delta^{2}\{S_{\mathrm{EM}}[\mathcal{A}]+\Gamma_{\psi\bar{\psi}}[\mathcal{A}]\}}{\delta\mathcal{A}^{\mu}(x)\delta\mathcal{A}^{\nu}(y)}A^{\mu}(x)A^{\nu}(y)+\cdots\bigg].
\end{align}
The resulting transition amplitude is therefore expressed as a coherent-field
path integral over histories connecting the asymptotic coherent-field
sectors  $\mathcal{A}_{(\mathrm{in})}$ and $\mathcal{A}_{(\mathrm{out})}$.
The first two terms in the action integral (\ref{eq: coherent state change in vacuum 03}),
\begin{align}
\Gamma_{\mathrm{HE}}[\mathcal{A}] & =S_{\mathrm{EM}}[\mathcal{A}]+\Gamma_{\psi\bar{\psi}}[\mathcal{A}],
\end{align}
denote the exact one-loop Heisenberg\textendash Euler effective action
for a fixed background field \citep{Dunne2005,Dittrich-Reuter1985}.
In practical calculations, this functional is often approximated by
the local Heisenberg\textendash Euler Lagrangian under the slowly
varying field approximation \citep{Karbstein2017}.

In contrast, the additional contribution $\Gamma_{A}[\mathcal{A}]$
originates from the functional integration over quantum photon fluctuations
around the coherent-field configuration and represents photon-loop
corrections beyond the conventional Heisenberg\textendash Euler theory.
Furthermore, the remaining functional integration 
\begin{equation}
\int_{\mathcal{A}_{(\mathrm{in})}}^{\mathcal{A}_{(\mathrm{out})}}\mathcal{D}\mathcal{A}
\end{equation}
sums over all coherent-field histories connecting distinct asymptotic
coherent-field sectors and thereby promotes the background field itself
to a dynamical transition variable. These structures are absent in
the conventional fixed-background Heisenberg\textendash Euler vacuum
model. 

We simplify the transition amplitude as
\begin{align}
\langle\alpha_{(\mathrm{out})\mathrm{H}}^{\prime},0_{(\mathrm{out})\mathrm{H}}^{A,\psi,\bar{\psi}}|\alpha_{(\mathrm{in})\mathrm{H}},0_{(\mathrm{in})\mathrm{H}}^{A,\psi,\bar{\psi}}\rangle & =N\int_{\mathcal{A}_{(\mathrm{in})}}^{\mathcal{A}_{(\mathrm{out})}}\mathcal{D}\mathcal{A}\exp\frac{i}{\hbar c}\Gamma[\mathcal{A}],\label{eq: coherent state change in vacuum 04}
\end{align}
with 
\begin{align}
\Gamma[\mathcal{A}] & =S_{\mathrm{EM}}[\mathcal{A}]+\Gamma_{\psi\bar{\psi}}[\mathcal{A}]+\Gamma_{A}[\mathcal{A}].\label{eq: Gamma mathcal A}
\end{align}
In the following effective-action analysis, we adopt a saddle-point
approximation to the coherent-field path integral \citep{Klauder-Skagerstam_1985,Negele-Orland1988}.
As discussed above, the notation $\int_{\mathcal{A}_{(\mathrm{in})}}^{\mathcal{A}_{(\mathrm{out})}}\mathcal{D}\mathcal{A}$
is a compact representation of endpoint coherent-state wave functionals
peaked around the asymptotic coherent fields. In the semiclassical
coherent-field regime considered below, the widths of these endpoint
wave functionals are neglected, so that the dominant contribution
may be described by a representative stationary coherent-field history
$\bar{\mathcal{A}}$. Within this approximation, the asymptotic coherent
fields are treated as effectively fixed boundary data for $\bar{\mathcal{A}}$.
We consider a stationary coherent-field configuration $\bar{\mathcal{A}}$
corresponding to $\Gamma[\mathcal{A}]$, satisfying
\begin{align}
\frac{\delta\Gamma[\mathcal{A}]}{\delta\mathcal{A}^{\nu}}\bigg|_{\mathcal{A}=\bar{\mathcal{A}}} & =0.\label{eq: Effective_Maxwell_eq_01}
\end{align}
We additionally impose the Lorenz gauge condition $\partial_{\mu}\bar{\mathcal{A}}^{\mu}(x)=0$
consistent with the Gupta\textendash Bleuler formalism. We decompose
the stationary coherent-field configuration $\mathcal{A}$ into $\bar{\mathcal{A}}$
and a fluctuation field $\delta\mathcal{A}$.
\begin{align}
\mathcal{A}(x) & =\bar{\mathcal{A}}(x)+\delta\mathcal{A}(x),
\end{align}
where the asymptotic boundary conditions are
\begin{align}
\bar{\mathcal{A}}(x)\rightarrow\mathcal{A}_{(\mathrm{in})}(x),\quad\delta\mathcal{A}(x)\rightarrow0,\quad & \mathrm{for}\quad t\rightarrow-\infty,
\end{align}
\begin{align}
\bar{\mathcal{A}}(x)\rightarrow\mathcal{A}_{(\mathrm{out})}(x),\quad\delta\mathcal{A}(x)\rightarrow0,\quad & \mathrm{for}\quad t\rightarrow+\infty.
\end{align}

The effective action $\Gamma[\mathcal{A}]$ can be evaluated around
the stationary coherent field $\bar{\mathcal{A}}$:
\begin{align}
\Gamma[\mathcal{A}] & =\Gamma[\bar{\mathcal{A}}]+\frac{1}{2}\int d^{4}x\int d^{4}y\frac{\delta^{2}\Gamma[\bar{\mathcal{A}}]}{\delta\bar{\mathcal{A}}^{\mu}(x)\delta\bar{\mathcal{A}}^{\nu}(y)}\delta\mathcal{A}^{\mu}(x)\delta\mathcal{A}^{\nu}(y)+O(\delta\mathcal{A}^{3}).
\end{align}
We neglect the terms of $O(\delta\mathcal{A}^{3})$ and higher in
the following discussion. For simplicity, we introduce
\begin{alignat}{1}
K_{\mu\nu}(x,y;\bar{\mathcal{A}}) & =\frac{\delta^{2}\Gamma[\bar{\mathcal{A}}]}{\delta\bar{\mathcal{A}}^{\mu}(x)\delta\bar{\mathcal{A}}^{\nu}(y)}=K_{\nu\mu}(y,x;\bar{\mathcal{A}}).\label{eq: Kernel_K}
\end{alignat}
The transition amplitude (\ref{eq: coherent state change in vacuum 04})
is approximated by the saddle-point expansion around the stationary
coherent-field configuration $\bar{\mathcal{A}}$,
\begin{align}
\langle & \alpha_{(\mathrm{out})\mathrm{H}}^{\prime},0_{(\mathrm{out})\mathrm{H}}^{A,\psi,\bar{\psi}}|\alpha_{(\mathrm{in})\mathrm{H}},0_{(\mathrm{in})\mathrm{H}}^{A,\psi,\bar{\psi}}\rangle\nonumber \\
 & =N\exp\frac{i}{\hbar c}\Gamma[\bar{\mathcal{A}}]\int_{\delta\mathcal{A}(t=-\infty)=0}^{\delta\mathcal{A}(t=+\infty)=0}\mathcal{D}\delta\mathcal{A}\exp\frac{i}{2\hbar c}\int d^{4}x\int d^{4}y\delta\mathcal{A}^{\mu}(x)K_{\mu\nu}(x,y;\bar{\mathcal{A}})\delta\mathcal{A}^{\nu}(y)\\
 & =N\exp\frac{i}{\hbar c}\bigg[\Gamma[\bar{\mathcal{A}}]+\frac{i\hbar c}{2}\mathrm{Tr}\ln K(\bar{\mathcal{A}})\bigg],\label{eq: laser depletion amplitude 01}
\end{align}
up to the standard normalization and regularization of the Gaussian
functional determinant \citep{Itzykson-Zuber1980,Negele-Orland1988}.
The effective action for the present coherent-field transition is
\begin{align}
\Gamma_{\mathrm{eff}}[\bar{\mathcal{A}}] & =\Gamma[\bar{\mathcal{A}}]+\frac{i\hbar c}{2}\mathrm{Tr}\ln K(\bar{\mathcal{A}}),\label{eq: Effective action}
\end{align}
where the term $(i\hbar c/2)\mathrm{Tr}\ln K(\bar{\mathcal{A}})$
arises from the Gaussian integration over fluctuations around the
stationary coherent-field configuration $\bar{\mathcal{A}}$.

\section{Discussion\label{sec:Discussion}}

The present formulation discussed in Sec.\,\ref{sec:Coherent_EM_transition}
should not be regarded as a replacement for the conventional Heisenberg\textendash Euler
framework \citep{Heisenberg-Euler1936,Schwinger1951,Dunne2005,Dittrich-Gies2000}.
Rather, it extends the standard background-field formulation by allowing
transitions between distinct coherent-field sectors \citep{Furry1951,Ritus1985,Piazza2012,Fedotov2023,Ilderton-Seipt2018}.
In the limit $\alpha^{\prime}=\alpha$, the asymptotic coherent-state
endpoint weights are centered on the same coherent-field configuration.
However, this condition alone does not recover the conventional fixed-background
formulation, since the functional integration over coherent-field
histories remains present. The conventional background-field description
is obtained only after further restricting the coherent-field path
integral to a single prescribed coherent-field history $\bar{\mathcal{A}}$.

\subsection{Effective Maxwell equation and induced polarization current}

We derived the equation of motion for the stationary coherent-field
configuration $\bar{\mathcal{A}}$ for the effective action $\Gamma_{\mathrm{eff}}[\bar{\mathcal{A}}]$
in Eq.\,(\ref{eq: Effective action}). Including the Gaussian fluctuation
determinant in the effective action, the fluctuation-corrected effective
Maxwell equation may be written as
\begin{align}
\frac{\delta\Gamma_{\mathrm{eff}}[\bar{\mathcal{A}}]}{\delta\bar{\mathcal{A}}_{\nu}(x)} & =\frac{1}{\mu_{0}}\partial_{\mu}\bar{\mathcal{F}}^{\mu\nu}(x)+\frac{\delta\Gamma_{\psi\bar{\psi}}[\bar{\mathcal{A}}]}{\delta\bar{\mathcal{A}}_{\nu}(x)}+\frac{\delta\Gamma_{A}[\bar{\mathcal{A}}]}{\delta\bar{\mathcal{A}}_{\nu}(x)}+\frac{i\hbar c}{2}\mathrm{Tr}\bigg[K^{-1}(\bar{\mathcal{A}})\frac{\delta K(\bar{\mathcal{A}})}{\delta\bar{\mathcal{A}}_{\nu}(x)}\bigg]=0
\end{align}
for $\bar{\mathcal{F}}^{\mu\nu}(x)=\partial^{\mu}\bar{\mathcal{A}}^{\nu}(x)-\partial^{\nu}\bar{\mathcal{A}}^{\mu}(x)$,
using the Lorenz-gauge condition:
\begin{align}
\partial_{\mu}\bar{\mathcal{F}}^{\mu\nu}(x)=\partial_{\mu}\partial^{\mu}\bar{\mathcal{A}}^{\nu}(x) & =\mu_{0}j_{\psi\bar{\psi}}^{\nu}(x)+\mu_{0}j_{A}^{\nu}(x)+\mu_{0}j_{\mathcal{A}}^{\nu}(x)
\end{align}
with the vacuum polarization current
\begin{align}
j_{\psi\bar{\psi}}^{\nu}(x) & =-\frac{\delta\Gamma_{\psi\bar{\psi}}[\bar{\mathcal{A}}]}{\delta\bar{\mathcal{A}}_{\nu}(x)},
\end{align}
the photon-fluctuation current
\begin{align}
j_{A}^{\nu}(x) & =-\frac{\delta\Gamma_{A}[\bar{\mathcal{A}}]}{\delta\bar{\mathcal{A}}_{\nu}(x)},
\end{align}
and the coherent-field (background laser field) fluctuation current
\begin{align}
j_{\mathcal{A}}^{\nu}(x) & =-\frac{i\hbar c}{2}\mathrm{Tr}\bigg[K^{-1}(\bar{\mathcal{A}})\frac{\delta K(\bar{\mathcal{A}})}{\delta\bar{\mathcal{A}}_{\nu}(x)}\bigg].
\end{align}
The effective current consists of the conventional vacuum-polarization
contribution $j_{\psi\bar{\psi}}^{\nu}(x)$ familiar from the Heisenberg\textendash Euler
theory \citep{Heisenberg-Euler1936,Schwinger1951,Dittrich-Gies2000,Bialynicka_Birula1970},
together with the additional contributions $j_{A}(x)$ and $j_{\mathcal{A}}(x)$
which arise from fluctuations around the coherent-state boundary conditions.
The latter encode the response of the coherent-field sector itself
and therefore provide a possible microscopic origin of coherent-field
depletion and backreaction. When $j_{A}(x),j_{\mathcal{A}}(x)\rightarrow0$,
the Maxwell equation $\partial_{\mu}\bar{\mathcal{F}}^{\mu\nu}(x)=\mu_{0}j_{\psi\bar{\psi}}^{\nu}(x)$
converges to the Heisenberg\textendash Euler vacuum model \citep{Euler_Kockel1935,Marklund2006,Dittrich-Gies2000},
i.e., the currents $j_{A}(x)$ and $j_{\mathcal{A}}(x)$ represent
corrections absent in the conventional fixed-background Heisenberg\textendash Euler
framework. They arise respectively from quantum-photon fluctuations
and coherent-field fluctuations, and therefore constitute candidate
contributions that may account for depletion and backreaction effects
of intense coherent laser fields.

The present formulation provides a systematic framework for evaluating
these additional contributions. In particular, the coherent-field
fluctuation current $j_{\mathcal{A}}(x)$ is formally determined by
the kernel $K(\bar{\mathcal{A}})$ through Eq.\,(\ref{eq: Kernel_K}).
Its explicit evaluation generally requires the photon propagator in
the coherent-field background and therefore involves the nonlocal
structure of $K(\bar{\mathcal{A}})$. A detailed quantitative analysis
of these fluctuation-induced currents is beyond the scope of the present
work and will be investigated elsewhere. Nevertheless, the effective
Maxwell equation derived above demonstrates that fluctuation-induced
corrections beyond the conventional Heisenberg\textendash Euler description
arise naturally within the coherent-field QED framework. These corrections
provide a possible route toward a systematic description of depletion
and backreaction effects.

\subsection{Depletion indicator}

We normalize the transition amplitude (\ref{eq: laser depletion amplitude 01})
by the vacuum persistence amplitude in the absence of a coherent field,
i.e.,
\begin{align}
\mathcal{M}{}_{\alpha\rightarrow\alpha^{\prime}} & =\frac{\langle\alpha_{(\mathrm{out})\mathrm{H}}^{\prime},0_{(\mathrm{out})\mathrm{H}}^{A,\psi,\bar{\psi}}|\alpha_{(\mathrm{in})\mathrm{H}},0_{(\mathrm{in})\mathrm{H}}^{A,\psi,\bar{\psi}}\rangle}{\langle0_{(\mathrm{out})\mathrm{H}}|0_{(\mathrm{in})\mathrm{H}}\rangle}\nonumber \\
 & =\exp\frac{i}{\hbar c}\bigg\{\Gamma[\bar{\mathcal{A}}]-\Gamma[0]+\frac{i\hbar c}{2}\mathrm{Tr}\ln[K(\bar{\mathcal{A}})K^{-1}(0)]\bigg\}.
\end{align}
The related transition probability within the saddle-point approximation
is
\begin{align}
P_{\alpha\rightarrow\alpha^{\prime}} & =|\mathcal{M}{}_{\alpha\rightarrow\alpha^{\prime}}|^{2}\nonumber \\
 & =\exp\bigg[-\frac{2\mathrm{Im}\{\Gamma[\bar{\mathcal{A}}]-\Gamma[0]\}}{\hbar c}-\mathrm{Re}\{\mathrm{Tr}\ln K(\bar{\mathcal{A}})K^{-1}(0)\}\bigg].\label{eq: transition probability of coherent states}
\end{align}
The present normalization removes field-independent determinant factors
and ensures $P_{0\rightarrow0}=1$. The transition probability $P_{\alpha\rightarrow\alpha^{\prime}}$
does not imply classical energy dissipation of a laser pulse during
free propagation. Rather, it characterizes the quantum probability
for transitions between coherent-field sectors induced by vacuum fluctuations.
The notion of depletion in the present work therefore refers to the
redistribution of probability among coherent-field sectors.

For $\alpha^{\prime}=\alpha$, the quantity $P_{\alpha\rightarrow\alpha}$
may be interpreted as the survival probability of the initial coherent-field
sector. Within the same saddle-point normalization, the complement
\begin{equation}
1-P_{\alpha\rightarrow\alpha}
\end{equation}
can be used as an indicator of the tendency of the system to leave
the original coherent-field configuration. In the present framework,
this quantity may be interpreted as a measure of coherent-field depletion
and backreaction induced by QED vacuum effects. Unlike conventional
fixed-background approaches, the coherent field itself is treated
as a dynamical quantum state, and transitions between different coherent-field
configurations can therefore be described explicitly through $P_{\alpha\rightarrow\alpha^{\prime}}$.
The conventional fixed-background approximation is recovered by prescribing
a single coherent-field history and excluding the integration over
transitions to coherent-field sectors with $\alpha^{\prime}\neq\alpha$.
Conversely, appreciable transition weight into such sectors provides
a signal of coherent-field depletion or backreaction beyond the fixed-background
approximation. 

This observation also clarifies the relation between the present formalism
and the conventional vacuum-persistence picture employed in strong-field
QED. In the latter approach, the probability loss
\begin{equation}
1-P_{\mathrm{vacuum}}(\bar{\mathcal{A}})
\end{equation}
is commonly associated with vacuum decay \citep{Schwinger1951} and
electron\textendash positron pair production \citep{Nikishov1970}.
More precisely, the quantity obtained from the imaginary part of the
effective action characterizes the vacuum-persistence probability,
which does not necessarily coincide with the mean pair-production
rate when multipair production becomes significant \citep{Cohen_McGady_2008}.
The vacuum-persistence probability does not resolve transitions between
different coherent-field sectors because the background field is treated
as fixed from the outset \citep{Ritus1985,Fedotov2023,GavrilovGitman1990}.
By contrast, the present formulation introduces the transition probability
$P_{\alpha\rightarrow\alpha^{\prime}}$ between distinct coherent
states and therefore provides a framework in which coherent-field
depletion may, in principle, be separated from real particle-production
channels.

This suggests that the reduction of the vacuum-persistence weight
conventionally described in a fixed-background treatment may, in a
more dynamical coherent-field description, contain contributions associated
with coherent-field transitions,
\begin{align}
\alpha\rightarrow\alpha^{\prime},\quad & \alpha^{\prime}\ne\alpha,
\end{align}
in addition to electron\textendash positron pair creation and other
quantum processes. This interpretation is analogous to the vacuum-persistence
probability introduced in strong-field QED \citep{Nikishov1970,Cohen_McGady_2008}.
The present coherent-field QED formulation offers a systematic starting
point for disentangling these contributions and investigating the
interplay between laser depletion, backreaction, and vacuum instability
within a unified QED framework.

\subsection{Weak EM coherent field estimate based on the Heisenberg\textendash Euler
effective action}

To gain qualitative insight into the depletion indicator given by
Eq.\,(\ref{eq: transition probability of coherent states}), we consider
the weak coherent-field limit, in which the EM coherent field is sufficiently
small compared with the QED critical field and the electron\textendash positron
contribution may be approximated by the leading-order Heisenberg\textendash Euler
effective action. The effective action introduced in the present formalism
is decomposed as
\begin{align*}
\Gamma[\bar{\mathcal{A}}] & =S_{\mathrm{EM}}[\bar{\mathcal{A}}]+\Gamma_{\psi\bar{\psi}}[\bar{\mathcal{A}}]+\Gamma_{A}[\bar{\mathcal{A}}],\tag{{\ref{eq: Gamma mathcal A}}}
\end{align*}
where $S_{\mathrm{EM}}[\bar{\mathcal{A}}]$ denotes the free EM field
action, $\Gamma_{\psi\bar{\psi}}[\bar{\mathcal{A}}]$ represents the
contribution arising from virtual electron\textendash positron fluctuations,
and $\Gamma_{A}[\bar{\mathcal{A}}]$ corresponds to radiative corrections
associated with photon fluctuations.

For the present estimate, we neglect the purely photonic contribution
$\Gamma_{A}[\bar{\mathcal{A}}]$ and retain only the leading vacuum-polarization
effect arising from virtual electron\textendash positron pairs. The
effective action is therefore approximated as
\begin{align}
\Gamma[\bar{\mathcal{A}}] & \simeq S_{\mathrm{EM}}[\bar{\mathcal{A}}]+\Gamma_{\psi\bar{\psi}}[\bar{\mathcal{A}}].\label{eq: Gamma_approx}
\end{align}
In the weak coherent-field limit, $\Gamma_{\psi\bar{\psi}}[\bar{\mathcal{A}}]$
is described by the leading-order Heisenberg\textendash Euler effective
action. Using the SI-unit expression \citep{Heisenberg-Euler1936,Euler_Kockel1935,Seto2014,Seto2015a,Seto2015b},
the nonlinear vacuum contribution is given by
\begin{align}
\Gamma_{\psi\bar{\psi}}[\bar{\mathcal{A}}] & =\int d^{4}x\frac{\alpha_{\mathrm{QED}}^{2}\hbar^{3}\varepsilon_{0}^{2}}{360m^{4}c}\{4[\bar{\mathcal{F}}_{\mu\nu}(x)\bar{\mathcal{F}}^{\mu\nu}(x)]^{2}+7[\bar{\mathcal{F}}_{\mu\nu}(x)\,{}^{*}\bar{\mathcal{F}}^{\mu\nu}(x)]^{2}\}
\end{align}
where $\bar{\mathcal{F}}^{\mu\nu}(x)=\partial^{\mu}\bar{\mathcal{A}}^{\nu}(x)-\partial^{\nu}\bar{\mathcal{A}}^{\mu}(x)$,
$^{*}\bar{\mathcal{F}}$ denotes its dual tensor, and $\alpha_{\mathrm{QED}}=e^{2}/4\pi\varepsilon_{0}\hbar c$
is the fine-structure constant. The fluctuation kernel introduced
in Eq.\,(\ref{eq: Kernel_K}) is defined by
\begin{align*}
K_{\mu\nu}(x,y;\bar{\mathcal{A}}) & =\frac{\delta^{2}\Gamma[\bar{\mathcal{A}}]}{\delta\bar{\mathcal{A}}^{\mu}(x)\delta\bar{\mathcal{A}}^{\nu}(y)}.\tag{{\ref{eq: Kernel_K}}}
\end{align*}
Under the approximation (\ref{eq: Gamma_approx}), it can be decomposed
as
\begin{align}
K_{\mu\nu}(x,y;\bar{\mathcal{A}}) & =K_{0\,\,\mu\nu}(x,y;\bar{\mathcal{A}})+\delta K{}_{\mu\nu}(x,y;\bar{\mathcal{A}}),
\end{align}
with
\begin{align}
K_{0\,\,\mu\nu}(x,y;\bar{\mathcal{A}})= & \frac{\delta^{2}S_{\mathrm{EM}}[\bar{\mathcal{A}}]}{\delta\bar{\mathcal{A}}^{\mu}(x)\delta\bar{\mathcal{A}}^{\nu}(y)},
\end{align}
\begin{align}
\delta K{}_{\mu\nu}(x,y;\bar{\mathcal{A}})= & \frac{\delta^{2}\Gamma_{\psi\bar{\psi}}[\bar{\mathcal{A}}]}{\delta\bar{\mathcal{A}}^{\mu}(x)\delta\bar{\mathcal{A}}^{\nu}(y)},
\end{align}
or in shorthand notation, 
\begin{align}
K & =K_{0}+\delta K.\label{eq: K-decomposition}
\end{align}
Since $\Gamma_{\psi\bar{\psi}}[\bar{\mathcal{A}}]$ begins at fourth
order in the field strength, its second functional derivative scales
as
\begin{align}
\delta K= & O(\bar{\mathcal{F}}^{2}).
\end{align}
Substituting Eq.\,(\ref{eq: K-decomposition}) into the determinant
term appearing in Eq.\,(\ref{eq: transition probability of coherent states}),
one obtains
\begin{align}
\mathrm{Tr}\ln(KK_{0}^{-1}) & =\mathrm{Tr}\ln(\mathbb{I}+K_{0}^{-1}\delta K).
\end{align}
For sufficiently weak coherent fields, $K_{0}^{-1}\delta K\ll\mathbb{I}$
and the logarithm may be expanded as
\begin{align}
\mathrm{Tr}\ln(KK_{0}^{-1}) & =\mathrm{Tr}(K_{0}^{-1}\delta K)+O(\bar{\mathcal{F}}^{4}).
\end{align}
The transition probability therefore becomes
\begin{align}
P_{\alpha\rightarrow\alpha^{\prime}} & =\exp\bigg[-\frac{2\mathrm{Im}\{\Gamma[\bar{\mathcal{A}}]-\Gamma[0]\}}{\hbar c}-\mathrm{Re}\{\mathrm{Tr}(K_{0}^{-1}\delta K)\}+O(\bar{\mathcal{F}}^{4})\bigg].\label{eq: transition probability in weak coherent field limit}
\end{align}
Equation (\ref{eq: transition probability in weak coherent field limit})
suggests that the corresponding coherent-field depletion indicator
is sensitive not only to the imaginary part of the effective action
but also to field-dependent modifications of the fluctuation spectrum
encoded in the kernel $K$. Within the present weak coherent-field
approximation, the leading contribution to this kernel modification
originates from electron\textendash positron vacuum polarization and
scales as $\delta K=O(\bar{\mathcal{F}}^{2})$. A quantitative evaluation
of the determinant term, as well as the inclusion of the purely photonic
contribution $\Gamma_{A}[\bar{\mathcal{A}}]$, is beyond the scope
of the present work and will be investigated elsewhere.

As a useful reference case, consider a single plane-wave coherent
field. For such a configuration, the EM invariants vanish identically,
$\bar{\mathcal{F}}_{\mu\nu}\bar{\mathcal{F}}^{\mu\nu}=0$ and $\bar{\mathcal{F}}_{\mu\nu}\,{}^{*}\bar{\mathcal{F}}^{\mu\nu}=0$.
It follows that all local Heisenberg\textendash Euler corrections
constructed from these invariants vanish, independently of the field
amplitude. The imaginary part of the effective action therefore satisfies

\begin{align}
\mathrm{Im}\{\Gamma[\bar{\mathcal{A}}]-\Gamma[0]\} & =0,
\end{align}
which is consistent with the well-known fact that a single plane wave
in vacuum does not induce Schwinger pair production or vacuum decay
\citep{Volkov1936,Ritus1985,Nikishov-Ritus1964,Nikishov-Ritus1967}.
Furthermore, within the local Heisenberg\textendash Euler approximation,
the fluctuation kernel is expected to remain close to its vacuum value,
\begin{align}
K(\bar{\mathcal{A}}) & \simeq K(0),
\end{align}
since the local vacuum-polarization corrections also vanish when the
EM invariants are zero. The exponent appearing in Eq.\,(\ref{eq: transition probability in weak coherent field limit})
is therefore strongly suppressed, implying that the coherent-field
history remains essentially unchanged within the present approximation.
The plane-wave case thus provides a useful consistency check of the
formalism. The absence of both vacuum decay and local Heisenberg\textendash Euler
corrections reproduces the standard result that an ideal prescribed
plane-wave background does not generate a nontrivial local vacuum
response. Nontrivial coherent-field transitions are therefore expected
to arise only for configurations with nonvanishing EM invariants,
nonlocal or derivative corrections, focusing effects, probe fields,
or superpositions of waves such as counterpropagating pulses.

As another example, consider a weak coherent-field configuration whose
electric component vanishes while the magnetic component remains nonzero,
\begin{align}
\bar{\boldsymbol{E}}=\boldsymbol{0},\quad & \bar{\boldsymbol{B}}\ne\boldsymbol{0},
\end{align}
as a formal weak-field background check. The detailed spatial profile
of the magnetic field is not important for the present discussion.
It may represent a uniform magnetic field, a slowly varying magnetic
configuration, or any weak coherent-field history for which the weak-field
expansion remains valid. In this case, the EM invariant $\bar{\mathcal{F}}_{\mu\nu}\,{}^{*}\bar{\mathcal{F}}^{\mu\nu}$
vanishes identically, whereas
\begin{align}
\bar{\mathcal{F}}_{\mu\nu}\bar{\mathcal{F}}^{\mu\nu} & =2\bigg(\bar{\boldsymbol{B}}^{2}-\frac{\bar{\boldsymbol{E}}^{2}}{c^{2}}\bigg)=2\bar{\boldsymbol{B}}^{2}
\end{align}
remains nonzero. Since no electric field is present, the conventional
Schwinger pair-production mechanism is absent and the effective action
contains no vacuum-decay contribution \citep{Greiner_Muller_Rafelski_1985},
\begin{align}
\mathrm{Im}\{\Gamma[\bar{\mathcal{A}}]-\Gamma[0]\} & =0.
\end{align}
The coherent-field transition probability therefore reduces to
\begin{align}
P_{\alpha\rightarrow\alpha^{\prime}} & =\exp[-\mathrm{Re}\{\mathrm{Tr}[K_{0}^{-1}\delta K]\}+O(\bar{\mathcal{F}}^{4})].
\end{align}
Thus, the leading contribution to the transition probability originates
not from vacuum decay but from the fluctuation determinant associated
with the coherent-field history itself. This example highlights a
conceptual difference between the present formulation and the conventional
Heisenberg\textendash Euler description. In the latter, a purely magnetic
background gives rise only to dispersive vacuum-polarization effects
and does not induce vacuum decay \citep{Heisenberg-Euler1936,Schwinger1951,Dittrich-Gies2000,Bialynicka_Birula1970}.
In contrast, within the coherent-field transition framework, modifications
of the fluctuation operator $K(\bar{\mathcal{A}})$ can still contribute
to transitions between distinct coherent-field sectors. The resulting
transition weight therefore provides a possible diagnostic of coherent-field
depletion and backreaction even in situations where pair production
is absent.

\section{Conclusion\label{sec:Conclusion}}

In this work, we have developed a coherent-field formulation of QED
in which laser fields are represented by coherent-state boundary conditions
rather than by prescribed classical backgrounds. Starting from the
Gupta\textendash Bleuler condition and the displacement-operator formalism,
we constructed both operator and path-integral descriptions of coherent-field
QED and derived transition amplitudes between distinct EM coherent
field configurations. While standard coherent-state path integrals
themselves are well known, the present formulation recasts them as
a coherent-field path integral, namely an endpoint-weighted description
of QED transitions between distinct EM coherent-field sectors specified
by asymptotic coherent states.

The resulting path-integral formulation leads naturally to an effective
action for the stationary coherent-field configuration. The corresponding
saddle-point condition yields an effective Maxwell equation containing
three distinct contributions: the conventional vacuum-polarization
current, a photon-fluctuation current, and a coherent-field fluctuation
current. In the limit where the latter two contributions vanish, the
formalism reduces to the conventional Heisenberg\textendash Euler
description. The present framework therefore extends the standard
fixed-background treatment by systematically incorporating quantum
fluctuations associated with the coherent field itself.

A second central result is the derivation of a saddle-point expression
for the coherent-field transition probability
\begin{equation}
P_{\alpha\rightarrow\alpha'}.
\end{equation}
For the special case $\alpha^{\prime}=\alpha$, the quantity $P_{\alpha\rightarrow\alpha}$
may be interpreted as the survival probability of a coherent-field
sector within the saddle-point normalization used here, while $1-P_{\alpha\rightarrow\alpha}$
provides a useful indicator of the tendency to leave the original
coherent-field sector. More generally, the transition probability
$P_{\alpha\rightarrow\alpha^{\prime}}$ provides a decomposition of
the conventional vacuum-persistence picture into transitions among
distinct coherent-field sectors. Since the standard fixed-background
formulation does not distinguish coherent-field transitions from other
quantum channels, the corresponding reduction in vacuum-persistence
weight may generally contain contributions associated with coherent-field
depletion in addition to particle-production processes. The present
formulation provides a framework in which these contributions can,
in principle, be identified and analyzed separately.

As a first qualitative application of the formalism, we also examined
the weak coherent-field limit using the leading-order Heisenberg\textendash Euler
effective action. Within this approximation, the depletion indicator
was shown to depend not only on the imaginary part of the effective
action but also on field-dependent modifications of the fluctuation
kernel associated with vacuum polarization. Although a quantitative
evaluation of these contributions remains for future work, this result
illustrates how coherent-field depletion can be analyzed within the
present effective-action framework.

It is also noteworthy that the present framework clarifies the conceptual
relation between coherent-field QED and conventional background-field
QED. Even when the asymptotic coherent states are chosen to be identical,
$\alpha^{\prime}=\alpha$, the transition amplitude generally still
contains a functional integral over coherent-field histories weighted
by the corresponding endpoint coherent-state overlaps. The conventional
background-field formulation (or the Furry picture) is recovered only
after imposing the additional restriction that a single coherent-field
configuration is prescribed throughout spacetime. From this perspective,
background-field QED may be viewed as a special limiting case of a
broader coherent-field transition formalism. This observation suggests
that coherent-field transitions contain information beyond that encoded
in the vacuum-persistence amplitude evaluated in a prescribed background
field. In this sense, coherent-field QED may be viewed as a framework
in which transitions between distinct coherent-field sectors are treated
as fundamental quantum processes.

The present work should therefore be regarded as a first step toward
a fully dynamical description of intense laser fields in coherent-field
QED. Future investigations will focus on explicit evaluations of the
fluctuation-induced currents and coherent-field transition probabilities
in realistic laser backgrounds. Particular attention will be devoted
to the quantitative determination of the fluctuation-determinant contribution
to the depletion indicator and to the role of photon-loop corrections
beyond the leading-order Heisenberg\textendash Euler approximation.

More broadly, the exploration of genuinely dynamical coherent-field
sectors, including their quantum backreaction and depletion dynamics,
represents an important direction for future investigations. The formalism
developed here provides a basis for future studies of coherent-field
depletion during ultra-intense laser pulse propagation, vacuum birefringence,
Schwinger pair production with dynamical laser backreaction, and scattering
processes involving transitions between distinct coherent-field sectors.

\section*{Acknowledgements}

This work was supported by the Japan Society for the Promotion of
Science (JSPS) Grant-in-Aid for Scientific Research No.\,24K06990
from MEXT of Japan.

\bibliographystyle{apsrev4-1}
\bibliography{SF-QED}

\end{document}